# Computational methods in cardiovascular mechanics

F. Auricchio (University of Pavia, Department of Civil Engineering and Architecture, auricchio@unipv.it)
M. Conti (University of Pavia, Department of Civil Engineering and Architecture, michele.conti@unipv.it)
A. Lefieux (Emory University, Division of Cardiology, adrien.lefieux@emory.edu)
S. Morganti (University of Pavia, Department of Electrical, Computer, and Biomedical Engineering, simone.morganti@unipv.it)
A. Reali (University of Pavia, Department of Civil Engineering and Architecture, alessandro.reali@unipv.it)
G. Rozza (SISSA, Mathematics, mathLab, Trieste, grozza@sissa.it)
A. Veneziani (Emory University, Department of Mathematics and Computer Science, avenez2@emory.edu, School of Advanced Studies IUSS Pavia)

## 1.0 Introduction: The Role and the Development of Computational Methods

The introduction of computational models in cardiovascular sciences has been progressively bringing new and unique tools for the investigation of the physiopathology. Together with the dramatic improvement of imaging and measuring devices on one side, and of computational architectures on the other one, mathematical and numerical models have provided a new – clearly noninvasive – approach for understanding not only basic mechanisms but also patient-specific conditions, and for supporting the design and the development of new therapeutic options. The terminology *in silico* is, nowadays, commonly accepted for indicating this new source of knowledge added to traditional *in vitro* and *in vivo* investigations. The advantages of *in silico* methodologies are basically the low cost in terms of infrastructures and facilities, the reduced invasiveness and, in general, the intrinsic predictive capabilities based on the use of mathematical models. The disadvantages are generally identified in the distance between the real cases and their virtual counterpart required by the conceptual modeling that can be detrimental for the reliability of numerical simulations.

In this respect, the terrific development of new devices and algorithms for image and data retrieval and processing allowed the migration from (over)simplified idealized descriptions to high-fidelity patient-specific models, in a critical – still ongoing – process of merging measures and conceptualizations. Meanwhile, the progressive improvement of computational resources provides the infrastructures to perform numerical simulations of complex dynamics in reasonable timelines. All such processes required, in background, the crucial development of novel specific modeling techniques and solvers in the field of computational mechanics, in an exciting process involving mathematics, engineering, computer science, and biomedical knowledge that is now having an impact not only on research but also on clinical practice. As a matter of fact, mathematical and computational modeling has been recognized more than a research tool, a service to be integrated in products for the clinical market, as the example of the company HeartFlow [Taylor et al., 2013] demonstrates.

The aim of this Chapter is to give an introduction to computational methods in cardiovascular mechanics. Far from being exhaustive, here we aim at providing basic concepts and examples for a minimal acquaintance with several references to the recent literature, possibly in view of a deeper investigation of the subject.

Accordingly, the first part of this Chapter is devoted to the numerical simulation of tissues and structures in cardiovascular mechanics. In particular, we deal with cardiovascular diseases and the structural simulation of the corresponding endovascular treatments, discussing the main steps and issues of these kinds of simulations. From the point of view of applications, we focus on one hand on the simulation of shape memory alloy stents and of carotid artery stenting, while on the other hand, we consider the important problem of the virtual simulation of transcatheter aortic valve implantation. Finite element analysis represents by far the most used simulation tool for the prediction of the structural behavior of tissues and devices and the simulation results that will be presented are obtained within this classical simulation framework. This section is however completed by a digression on a very promising extension of finite elements known as Isogeometric Analysis.

In the second part of the Chapter, we aim at providing the reader a "survival kit" for entering the fascinating world of numerical modeling of the human hemodynamics. As a matter of fact, the complexity of the systems reflects in a significant complexity of the mathematical and numerical problems to tackle. We can state that computational hemodynamics has provided in the last twenty years a practical framework for many theoretical and methodological developments relevant for a much wider range of applications (in fact, this is true for the celebrated Euler equations, introduced by the Swiss Mathematicians for describing blood flow in compliant arteries and eventually used for describing the gas dynamics in pipes like in internal combustion engines). In particular, we will consider here basic concepts of numerical modeling of blood as a fluid in 3D domains.

We will then complement these first two parts of the Chapter with a discussion on specific methods to simulate the complexities that may arise when considering fluid-structure interaction problems occurring in computational hemodynamics both for the interaction between blood and the vascular walls as well as for the valves dynamics. Moreover, as the computational costs required by the numerical approximation of the considered problems is generally high, their complexity may need to be conveniently reduced, calling for specific model reduction techniques based on the online/off-line paradigm. Those techniques take advantage from previous available high-fidelity simulations to perform rapid computations on different cases. We will, therefore, complete the final part of this Chapter giving a primer on reduced order model methods, looking particularly promising to fulfill clinical timelines.

## 2.0 Simulating Tissues and Structures in Cardiovascular Mechanics

### Cardiovascular diseases and endovascular treatments

Cardiovascular disease (CVD) is the generic name given to the dysfunctions of the cardiovascular system such as atherosclerosis, hypertension, coronary heart disease, heart failure, or stroke. CVD is still the main cause of death in Europe, leading to almost two times as many deaths as cancer across the continent [Townsend et al, 2015]. In particular, within the broad family of CVD, we will refer in the following to focal obstructive lesions - stenosis - of the arteries (coronaries, carotid, and limb arteries) and heart valves, or abnormal localized bulging of the aorta, i.e., aneurysm. For such a class of vascular diseases, historically treated by combining open surgery with medical management, the use of endovascular approaches has revolutionized the treatment of vascular disease. In fact, in recent decades, endovascular therapy of vascular diseases has expanded from simple dilatation of atherosclerotic lesions to more complex acute lesions, such as aortic aneurysms or dissections. As already stated, these broadened indications for endovascular therapy have been supported by improvements in the concept, design, and technological content of endovascular devices. Such an advancement has been supported by dedicated biomechanical analysis of the artery-device interactions through computational tools, like structural Finite Element Analysis or Computational Fluid Dynamics, which are nowadays extensively used during the device design [Alaimo et al., 2017], for pre-operative planning [Morganti et al., 2016, de Jaegere et al., 2016], or diagnostics [Gasser et al., 2016, Gaur et al., 2017] as discussed in the following, where we deal with the different aspects of simulating tissues and structures in cardiovascular mechanics. In particular, we focus here on the simulation of endovascular treatments of peripheral arteries (e.g., carotid artery) and aortic valve, neglecting coronary stenting, which would deserve a dedicated dissertation as reported in [Morlacchi et al., 2013].

## 2.1 Simulation framework: from medical images to the virtual endovascular implant

Over recent years, numerical methods and scanning technology such as Computed Tomography (CT) and Magnetic Resonance (MRI) have advanced rapidly, calling for quality mesh generation, directly from images, enabling the analysis of complex biomedical phenomena. Common approaches to generate vascular meshes extract boundary surfaces using isocontouring [Yushkevich et al., 2006], which usually involves manual interaction, and then construct tetrahedral [Antiga et al., 2008] or hexahedral (hex) meshes [De Santis et al., 2011, Bols et al., 2016]. Dedicated algorithms have been proposed [Zhang et al., 2007] to generate hexahedral solid NURBS (Non-Uniform Rational B-Splines) meshes for patient-specific vascular geometric models from imaging data for use in isogeometric analysis [Hughes et al., 2005] as well.

The direct assessment from CT and MRI images of arterial wall thickness, which is an important part of the computational modeling of the vessel wall, is a non-trivial task in most cases, due to the limitation of imaging resolution. A number of approaches have been proposed in the literature. Johnson and colleagues [Johnson et al., 2011] deformed a healthy vessel onto a cerebral aneurysm by surface parameterization and a nonlinear spring to construct the corresponding model of weakened wall, estimating the material strength and anisotropy (e.g., an equivalent wall thickness and the different Young's moduli) by comparing the original surface and the deformed mesh [Zhang et al., 2013]. The use of such an equivalent wall thickness seems to yield a more accurate prediction of the aneurysm rupture site. Anyway, in the case of mechanical simulations of aortic aneurysms, the assessment of wall thickness is still an open issue [Gasser, 2016], and in particular the assumption of uniform wall strength and thickness in FE models predicting aneurysm rupture seems to provide better results than using a variable wall thickness [Martufi et al., 2015]. As a matter of fact, most of the structural simulations dealing with the analysis of the deployment of endovascular devices and the consequent interaction with the arterial wall assume either an elastic wall with uniform thickness [Altnji et al., 2015, Perrin et al., 2015] or a wall represented as a rigid surface [Auricchio et al., 2013a, Cosentino et al., 2015]. Instead, the inclusion of calcifications and an accurate modeling of mechanical non homogeneity due to atherosclerotic degeneration of arterial wall still imply manual segmentation [Auricchio et al., 2013b] and are limited to a small population of patients [Wang et al., 2017]. Readers interested in these topics are referred to the contribution of Zhang et al. [Zhang et al., 2016], reporting extensive information about geometric modeling and mesh generation from scanned images in the biomedical field.

**Boundary conditions**

Mechanical loading conditions experienced by the endovascular devices implanted into the cardiovascular system can be complex, potentially impairing the performance and the mechanical durability of the device. Although cyclic radial expansion of the arteries, due to the pulsatile action of the luminal blood pressure, has been the main focus of cardiovascular device durability in the past [Pelton et al., 2008], nowadays we acknowledge the importance of other loading components due to arterial stretching, shortening, bending, twisting, and kinking, which are induced not only by the cardiac and the respiration cycle [Ullery et al., 2015] but also by the musculoskeletal movement, e.g., leg bending [MacTaggart et al., 2014] or swallowing and neck twisting [Robertson et al., 2008]. Such considerations are particularly remarkable for specific vascular districts such as Superficial Femoral Arteries (SFA) of popliteal artery, which are close to important musculoskeletal joints (i.e., hip and knee) and at the same time bounded by muscular tissue. Indeed, SFA biomechanics has been recently investigated by advanced medical imaging [Cheng et al., 2006] or by ex-vivo experiments [Poulson et al., 2017] with the final aim to assess the loadings induced to the implanted stents [Choi et al., 2009], imposing directly the measured arterial kinematics [Conti et al., 2017], or to calibrate arterial model parameters [Petrini et al., 2016]. The situation is further complicated by the age-dependency of the arterial motion [Kamenskiy et al., 2015] and tortuosity [Thomas et al., 2015]. Moreover, biomechanical asymmetry is widely diffused in the body, as proved by the in-vivo measurement of the renal artery deformation [Suh et al., 2013]. Finally, the presence of implanted devices can dramatically change the dynamic vascular environment [Hirotsu et al., 2017; Nauta et al., 2017], inducing also a stiffening of the whole arterial district [de Beaufort et al., 2017a, de Beaufort et al., 2017b]. This suggests that the sole pre-operative data are not enough to cover the whole range of loading that the device will experience. Given such considerations, it is evident that the assessment of dedicated boundary conditions in the biomechanical analysis of endovascular devices is fundamental, calling for patient-site-specific investigation. In this respect, there is also a gap of knowledge that has to be filled, from both academia and industry, with the final common aim to design, benchmark, and manufacture long-lasting implanted endovascular devices such stents or endografts.

**Device modeling: SMA-based stent simulations**

Thanks to its unique mechanical features such as pseudo-elasticity, Nitinol (the most well-known and used shape memory alloy, made of the combination of Nickel and Titanium) has allowed the design of many innovative applications in the biomedical field [Auricchio et al., 2015]. Among the others, cardiovascular self-expanding stents is the most successful case of Nitinol, having an important commercial interest and calling for engineering tools supporting the design of novel devices. To this aim, during the device design stage, it is of paramount importance to accurately predict the complex behavior of such a shape memory alloy (SMA), under various loading conditions by dedicated constitutive modeling. The pioneering study of Witcher in 1997 [Whitcher, 1997] used structural finite element analysis (FEA) to estimate mechanical behavior of *Boston Scientific*'s *Symphony* stent under in-vivo loading conditions. Several simplistic assumptions are present: a von Mises-yield elasto-plastic constitutive law to model Nitinol, while pressure load on a stent portion resembles the in-vivo loading conditions neglecting the inclusion of the arterial model. In 2000, Rebelo and Perry [Rebelo, Perry, 2000] adopted the constitutive model by Auricchio and Taylor [Auricchio, Taylor, 1997, Auricchio et al., 1997] based on the concept of generalized plasticity [Lubliner, Auricchio, 1996] to take into account pseudo-elasticity in the FEA simulation of Nitinol stent self-expansion. In 2002, Perry and colleagues [Perry et al., 2002] have extended the study to analyze stent fatigue resistance, which was further investigated in 2003 by Pelton and colleagues [Pelton et al., 2003], who combined displacement-controlled fatigue experimental tests on laser-cut stent-like devices with non-linear FEA to compute strains. Furthermore, in 2004 Auricchio and Petrini [Auricchio, Petrini, 2004] proposed a very robust SMA constitutive model able to reproduce both pseudo-elastic and shape-memory effects, and particularly suited for the simulation of real industrial applications, like the design of medical stents [Petrini et al., 2005]. In 2006, Thèriault and colleagues [Thèriault et al., 2006] discussed the development of a Nitinol stent able to smoothly expand in the artery wall by creep effect of a polymeric cover. In 2008, Kim et al. [Kim et al., 2008] discussed the mechanical modeling of self-expandable braided stents, proposing a finite element model coupled with the preprocessing program for the three-dimensional geometrical modeling of the braiding structure. In 2010, Auricchio et al. [Auricchio et al., 2010] reviewed the properties of the SMA three-dimensional model described in [Auricchio, Petrini, 2004], calibrating the model parameters with respect to experimental data, and showing its application to the simulation of pseudo-elastic Nitinol stent deployment in a simplified atherosclerotic artery model. In 2011, Rebelo and colleagues [Rebelo et al., 2011] presented a study in which some commonly made assumptions in FEA of Nitinol devices were verified. In 2012, Garcia and colleagues [Garcia et al., 2012] using FEA performed a parametric analysis of a commercial stent model to estimate the influence of geometrical variables on the stent radial expansion force. Other recent studies target the stent design optimization using FEA such as Hsiao and Yin [Hsiao et al., 2013] who proposed to shift the highly concentrated stresses/strains away from the stent crown and re-distribute them along the stress-free bar arm by tapering its strut width. Alaimo et al. [Alaimo et al., 2017] have conformed such a design approach extending to a multi-objective optimization framework the approach proposed by Azaouzi and colleagues [Azaouzi et al., 2013].

**Carotid artery stenting**

Endovascular treatment of carotid stenosis represents one of the main fields of vascular surgery, supported by a strong industrial interest: in fact, the global neurovascular intervention market is expected to surpass \$2.5 billion by 2018. Such an interest leads to continuous development of novel devices [Schofer et al., 2015], stimulating also the biomechanical studies addressing the use of structural FEA to design or optimize the carotid stent design. The first study in this direction was done in 2007 by Wu et al. [Wu et al., 2007], who simulated the implant of a Nitinol stent in a geometrical idealization of the carotid bifurcation, accounting for the delivery sheath. Following the study of Wu, in 2011, Auricchio et al. [Auricchio et al., 2011] used structural FEA to evaluate the performance of three different self-expanding stent designs in the same carotid artery model, based on computed angiography tomography images, towards a quantitative assessment of the relationship between a given carotid stent design and a given patient-specific carotid artery anatomy. The same computational framework was subsequently validated in vitro [Conti et al., 2011] and used to assess the impact of stent scaffolding [Auricchio et al., 2012], as well as the impact of plaque morphology and anisotropic behavior in the arterial modeling [Auricchio et al., 2013b]. More recently, Iannaccone et al. [Iannaccone et al., 2014] investigated the role of plaque shape and composition on the carotid arterial wall stress after stenting, using structural finite element simulations and generalized atherosclerotic carotid geometries including a damage model to quantify the injury of the vessel. An example of the computational framework aiming at simulating the carotid stent deployment in a patient-specific model derived from medical images is illustrated in MC_CAS_Image Figure 1.

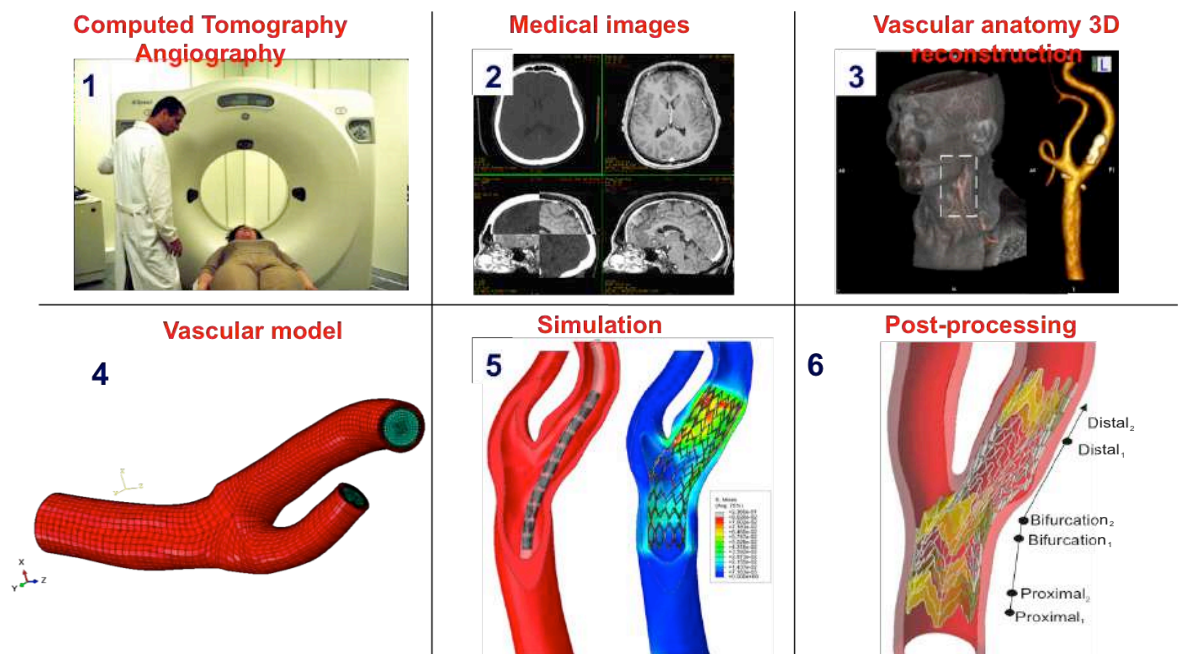

Figure 1. Illustrative representation of the steps to be performed for the patient-specific simulation of an endovascular procedure (in this case, carotid artery stenting). Starting from medical images routinely used in clinical practice, an accurate 3D model of the vascular anatomy can be created and transformed in an analysis-suitable domain. Combining the vascular model with a given prosthesis model is possible to virtually predict the post-implant vessel/prosthesis configuration and compute, through a dedicated post-processing, clinically-relevant measures (almost impossible to quantify in vitro or in vivo).

**Transcatheter aortic valve implant (TAVI)**

In the last decade, computational tools have been increasingly and extensively used for the virtual simulation of transcatheter aortic valve implantation (TAVI). The reason is twofold: on one side, from the medical point of view, TAVI is turning out to be not only a consolidated minimally-invasive technique for inoperable patients but also a very promising solution even in high or intermediate risk patients [Smith et al., 2010]; on the other side, from the engineering point of view, computational tools and simulation technologies are becoming more and more powerful, allowing realistic virtual reproduction of real, even complex, procedures in short time. Analogously to other cardiovascular computational models, in order to satisfy accuracy and reliability requirements, TAVI models have to take into account the patient-specific anatomical details, the characteristics of the patient's arteries (e.g., appropriate constitutive models), the boundary conditions and the loads governing the prosthesis expansion, as well as those generated by the procedure acting on the anatomical structures. From the first publication by Dwyer et al. [Dwyer H.A., et al., 2009] aiming at characterizing the blood ejection force able to induce a prosthesis migration, several other works have been published also using patient-specific data. We here recall the first study using patient's data (from a 68-year-old male), proposed by Sirois et al. [Sirois et al., 2011]. The procedure of TAVI is quite complex and its main steps can be summarized in (a) device crimping, (b) positioning, and (c) expansion. Additionally, each of the previously listed steps involves non-trivial physical phenomena, as the strong crimping of a complex-shape device made of materials exhibiting non-linear behavior or the expansion of the same device that interacts with native tissues and calcifications. For this reason, many authors have focused their work on specific aspects of the entire TAVI procedure. Wang et al. [Wang et al., 2012], for example, focused only on the deployment of a balloon-expandable device within a patient-specific aortic root reconstructed from medical images. Analogously, Gunning et al. [Gunning et al., 2014] analyzed in a patient-specific case the bioprosthetic leaflet deformation due to the deployment of a self-expanding valve. For simplicity, some authors considered only the stent for their numerical investigations: Schievano et al. [Schievano et al., 2010] and Capelli et al. [Capelli et al., 2014] proposed a FEA-based methodology to provide information and help clinicians during percutaneous pulmonary valve implantation planning. In these works, the implantation site has been simplified using rigid elements and, at the same time, the presence of the valve has been neglected. On the contrary, many other studies focus on the leaflets neglecting the stent: for example, Smuts et al. [Smuts et al., 2011] developed new concepts for different percutaneous aortic leaflet geometries by means of FEA, while Sun et al. [Smuts et al., 2010] have investigated the implications of asymmetric transcatheter prosthesis deployment on the bioprosthetic valve. Capelli et al. [Capelli et al., 2012] performed patient-specific analyses to explore the feasibility of TAVI in morphologies which are borderline cases for a minimally-invasive approach. Tzamtzis et al. [Tzamtzis et al., 2013] compared the radial force produced by a self-expandable vale (i.e., the Medtronic Corevalve) and a balloon-expandable one (i.e., the Edwards Sapien valve). The radial force of a self-expanding valve was also investigated by Gessat et al. [Gessat et al., 2014] who developed an innovative method for extracting such a force measure from images of an implanted device. It is worth mentioning the other following works by Auricchio et al. [Auricchio et al., 2014] and Morganti et al. [Morganti et al., 2014, Morganti et al., 2016] that proposed a step by step strategy to simulate the entire implantation procedure (from crimping to expansion) of both balloon-expandable and self-expandable prosthetic valves.

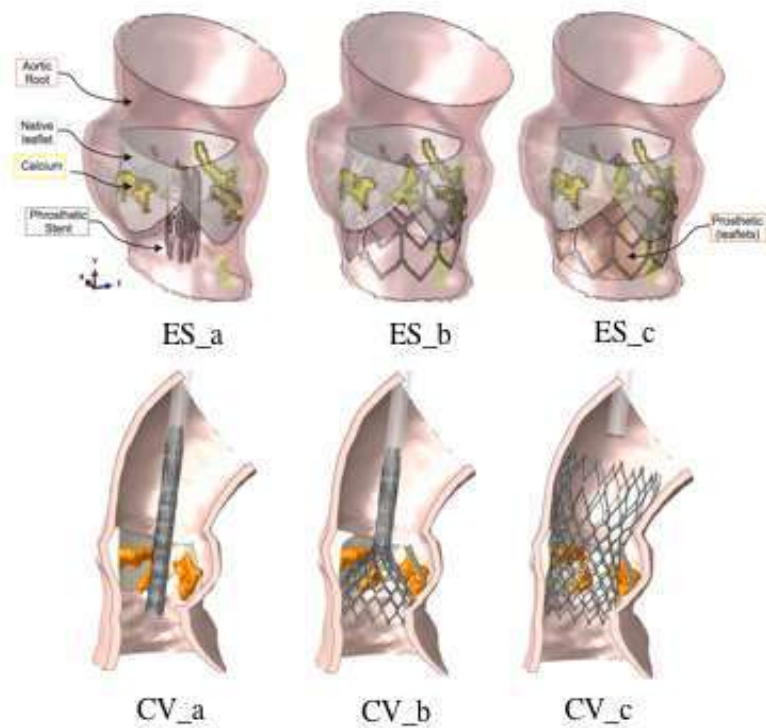

Figure 2. TAVI simulation steps: Edwards Sapien XT stent positioning (ES_a), device expansion (ES_b), post-implant valve performance (ES_c). Medtronic Corevalve stent positioning (CV_a), Device opening (CV_b),  and implanted device final configuration (CV_c).

The developed simulation frameworks, based on patient-specific imaging data, allow predicting very useful clinical parameters, proposing computational tools as a promising support for the decision-making process. In particular, through predictive simulations, it is possible to know in advance the following results that may be correlated with possible procedure complications. The impact of the metallic frame of the stent on the native calcified aortic root wall can be evaluated for example by computing the von Mises aortic wall stresses induced by stent expansion. On one side, higher stress values can be related to higher force of adherence between stent and aortic wall; on the other side, high-stress patterns concentrated in the annular region can indicate a major risk of aortic rupture. Besides the magnitude of the force induced by the device on the aortic wall, the grade of prosthesis apposition and, consequently, a correspondent measure of device anchoring could be evaluated by measuring the area of the contact surface between the stent and the aortic root. At the same time, structural FEA can be used to quantitatively evaluate the area of perivalvular orifices, which can be assumed to be proportional to the amount of retrograde perivalvular blood flow (i.e., perivalvular leakage). Moreover, the native morphology of the aortic root and, in particular, the quantity and position of calcifications may induce a non-circular shape to the implanted device that can impact on the valve performance and can be predicted by finite element analysis. From the simulation of valve closure, we can finally predict the post-operative device performance in terms of valve coaptation.

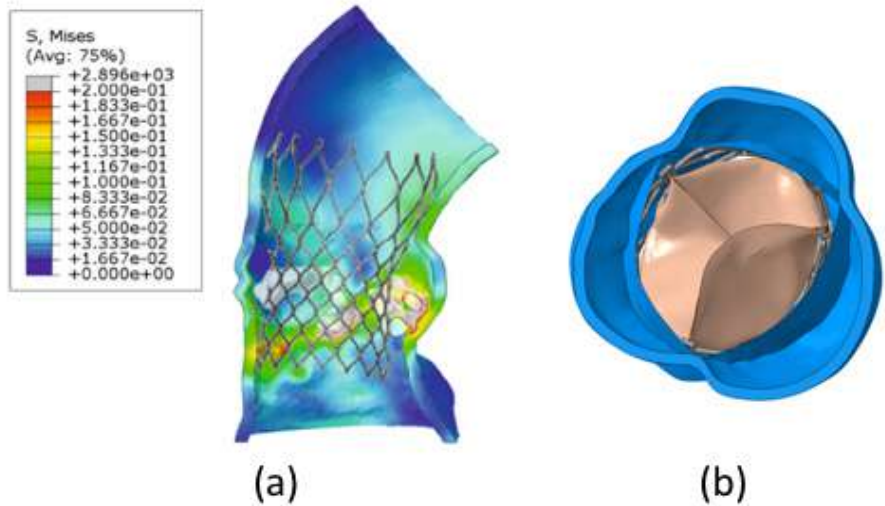

**Beyond classical finite element methods: Isogeometric Analysis**

Finite element analysis represents by far the most used simulation tool for the prediction of the structural behavior of tissues and devices. However, the research community is continuously working to develop evolutions of classical FEA or completely new concepts able to go beyond the limits of standard finite elements in terms of simulation speed and accuracy. A complete survey of alternative approaches is beyond the scope of this chapter and we limit the discussion here to IsoGeometric Analysis (IGA, see [Cottrell et al., 2009]), certainly one of the most celebrated (relatively) new options to enhance FEA performance.

IGA is a relatively novel computational technology, introduced in 2005 by Hughes and coworkers [Hughes et al., 2005] with the main aim of bridging the gap between Computer Aided Design (CAD) and the FEM-based engineering analysis process. Its basic paradigm consists of adopting the same basis functions used for geometry representations in CAD systems - such as, e.g., Non-Uniform Rational B-Splines (NURBS) - for the approximation of field variables, in an isoparametric fashion. This leads to a cost-saving simplification of the typically expensive mesh generation and refinement processes required by standard FEA, which was the original motivation for IGA. Moreover, thanks to the high-regularity properties of its basis functions, IGA has shown a better accuracy per-degree-of-freedom and an enhanced robustness with respect to standard FEA in a number of applications. Solids and structures are a prime example (see, e.g., [Cottrell et al., 2006; Cottrell et al., 2007; Elguedj et al., 2008; Lipton et al., 2010; Schillinger et al., 2012; Caseiro et al., 2015]), including also effective beam, plate, and shell elements (see, e.g., [Kiendl et al., 2009; Benson et al., 2010; Echter et al., 2013; Kiendl et al., 2015]). IGA has moreover been successful in fluid mechanics and fluid-structure interaction (see, e.g., [Bazilevs et al., 2007; Akkerman et al., 2008; Gomez et al., 2010; Hsu, Bazilevs, 2012; Hsu et al., 2015]), and has opened the door to geometrically flexible discretizations of higher-order partial differential equations in primal form (see, e.g., [Auricchio et al., 2007; Gomez et al., 2008; Kiendl et al., 2016]). Thanks to its more than promising results, IGA attracted a tremendous deal of attention and is now regarded as one of the most prominent research areas of modern Computational Mechanics. Within this context, we focus here on two recent applications able to show the potential of IGA for structural biomechanical simulations, dealing with aortic valves and SMA stents.

The first study we mention is related to explicit dynamics simulations of the closure of a patient-specific aortic valve, which have been performed via IGA using LS-DYNA [Morganti et al., 2015]. The adopted complex geometrical model has been built, starting from medical images, by means of conforming multi-patch untrimmed NURBS and, on such a model, nonlinear shell analyses involving large deformations and contact have been successfully performed. Mesh convergence studies targeting a good approximation of coaptation. Despite the lack of optimization of the adopted IGA implementation, for a given target level of accuracy, the IGA simulation has resulted to be two orders of magnitude faster than that performed with what is considered to be the fastest shell finite element on the market, indeed a very remarkable result.

Qualitatively similar results have been carried out also in the context of the nonlinear static analysis of SMA stent structures by means of 3D solid elements [Auricchio et al., 2015]. It was clearly confirmed there that, when dealing with complex nonlinear phenomena (these simulations include large deformations, complex inelastic constitutive laws, and buckling), standard low-order FEA may fail in correctly reproducing the underlying physics of the problem unless using extremely (sometimes excessively) fine meshes, while IGA can correctly tackle it with (relatively) very coarse meshes. The gain in terms of computational time with respect to FEA has been shown again to be very significant also in these static simulations, where, for the same accuracy, IGA is over one order of magnitude faster than FEA.

## 3.0 Simulating Fluids in Cardiovascular Mechanics

The nature of blood as a fluid was already elucidated in the literature [Nichols et al., 2011, Pedley, 1980, Formaggia et al., 2009, Galdi et al., 2008] and in Chapter 4. Here we recall the basic features relevant for the selection of a specific mathematical and computational model.

Blood is a complex suspension in an aqueous solution (the plasma) of several particles, including red cells, blood cells, and platelets. This nature clearly affects the physical behavior of blood as a liquid and, eventually, the selection of a mathematical model for blood flow. A decisive, extrinsic aspect in this is the domain where blood flows, which is significantly heterogeneous, ranging from a large vessel like the aorta to a huge number of small capillaries. The relative size of the vessel compared to the size of the particles convected by the bloodstream is critical in providing an appropriate mathematical description of the rheology, i.e., the constitutive law describing the internal actions of the fluid as a continuum. As a matter of fact, a continuum description of blood flow in the capillaries may be inappropriate as opposed to a particle modeling, since the cells in the plasma have dimensions comparable with the size of the vessels. The region of interest of the vascular system is important also for the flow regime. While in large and medium-size vessels, blood flow features a significant unsteadiness or, more precisely, *pulsatility* [Nichols et al., 2011], with a prevalence of convective forces, the small vessels and the capillary bed feature typically a quasi-steady dynamics, dominated by viscous forces. In the latter case, flow is clearly laminar, while in large vessels we may observe physical disturbances due to the convective forces. While in some animals this flow may show turbulent dynamics, in human beings this is occurring in some pathological conditions, since the sequencing of a strong acceleration (lasting one-third of the heartbeat) during the so-called *systole*, followed by a relatively quiet phase called *diastole* (when the heart valve is closed) normally dumps the transition to turbulence.

In this Chapter, we specifically focus on blood flow in large and medium arteries. In fact, these are generally the sites of the most important vascular pathologies. We point out, however, that we will cover numerical aspects of computational hemodynamics that go beyond the specific realm of large arteries. Specifically, we will consider blood as

1) Incompressible, so that density $\rho$ is constant;
2) Newtonian, so that the rheology is simply described by a linear relation between the strain and the stress tensors, the proportionality coefficient $\mu$ (dynamic viscosity) being a constant;
3) Pulsatile, so that we account for inertial forces;
4) Laminar, so that no specific modeling of turbulence is required. We will cover briefly the numerical treatment of turbulence, though, as this is crucial for some vascular districts.

These are the usual modeling simplifying assumptions for the description of blood flow in large arteries, and they lead to the celebrated system of *the incompressible Navier-Stokes equations* to be the basic core for computational hemodynamics in 3D.

Denoting by $u(x,t)$ the velocity vector and by $p(x,t)$ the pressure field, function of the space vector-variables $x$ and time $t$, the incompressible Navier-Stokes Equations (NSE) in a region of interest $\Omega$, like the one illustrated in Figure 4, read

$$\begin{cases} \rho(\partial_t \boldsymbol{u} + (\boldsymbol{u} \cdot \nabla)\boldsymbol{u}) - \mu \nabla \cdot (\nabla \boldsymbol{u} + \nabla \boldsymbol{u}^{\mathrm{T}}) + \nabla p = \boldsymbol{f}, \\ \nabla \cdot \boldsymbol{u} = 0 \end{cases} \tag{1}$$

for $\boldsymbol{x} \in \Omega$ and $t > 0$, where $\boldsymbol{f}$ is a (generic) forcing term (e.g., the gravity).

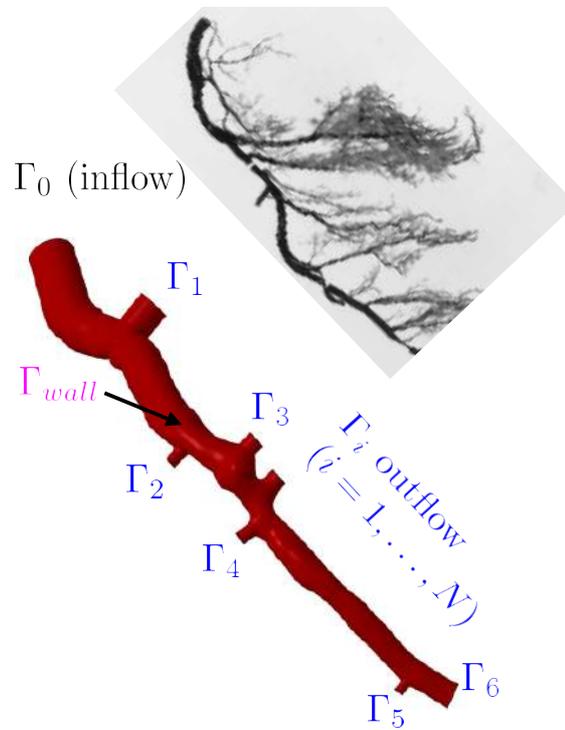

$\Gamma_0$ (inflow)

$\Gamma_1$

$\Gamma_{wall}$

$\Gamma_3$

$\Gamma_2$

$\Gamma_i$ outflow ($i = 1, \ldots, N$)

$\Gamma_4$

$\Gamma_6$

$\Gamma_5$

Figure 4. Representation of a typical domain in computational hemodynamics (a coronary artery). The different portions of the boundary are highlighted.

Here, the first equation represents the momentum conservation for a Newtonian fluid and the second one follows from the mass conservation.

The problem describing the blood dynamics needs to be completed by *initial conditions* on the velocity (the only unknown under time derivative) and by *conditions to prescribe on the boundary* of $\Omega$ that we will denote by $\Gamma$. These conditions represent the mutual influence on the blood by the external tissues and the upstream/downstream circulation. Generically, the initial conditions are prescribed in the form

$$\boldsymbol{u}(\boldsymbol{x}, 0) = \boldsymbol{u}_0(\boldsymbol{x}) \quad \forall x \in \Omega$$

where $\boldsymbol{u}_0(x)$ is assumed to be given. However, an accurate knowledge of the velocity field in the entire domain of interest is difficult to retrieve from current measurement devices. In practice, it is a good simulation practice to prescribe an arbitrary velocity field and then to exploit the periodicity of the blood flow, so to compute a certain number of heart beats. Typically, after a certain number of heart beats (in the range of 3-10, depending on the vascular district at hand), the influence of the initial condition on the numerical solution is significantly lost.

The issue of the boundary conditions is much more delicate, as they have a major impact on the computed solution. Unfortunately, in most of the cases, a complete data set to prescribe boundary conditions is not available in practice and the appropriate identification of numerical strategies to fill the gap between the mathematical theory of the NSE and simulation practice is still subject of ongoing research.

### 3.1 Boundary conditions: Theory and Practice

From the mathematical point of view, generally speaking, the boundary conditions mostly occurring in the simulation of blood flow are of two types

- *velocity conditions*, i.e., the prescription of the velocity field $\boldsymbol{v}(\boldsymbol{x}, t)$ in every point of the boundary

$$\boldsymbol{u}(\boldsymbol{x}, t) = \boldsymbol{v}(\boldsymbol{x}, t) \text{ for } \in \Gamma; \tag{2}$$

- *traction conditions*, i.e., conditions prescribing the normal component of the stress tensor occurring in the momentum equation, in the form

$$p\boldsymbol{n} - \nu(\nabla \boldsymbol{u} + \nabla^T \boldsymbol{u}) \cdot \boldsymbol{n} = \boldsymbol{d}, \tag{3}$$

where $\boldsymbol{d}$ is given.

These conditions are both vector data, enforcing three scalar functions in each boundary point. They are mathematically "correct". With this, we mean that they correctly complete the Navier-Stokes equations set, and under some additional assumptions on the regularity of the domain $\Omega$ and on the initial conditions, it is possible to prove that the problem has a unique solution (technically, we say that the problem is *well posed*) [Temam, 1984, Temam, 1995]. Mathematically speaking, velocity conditions are usually called "Dirichlet" conditions, while traction data are also called "Neumann" conditions.

Unfortunately, in the simulation of blood flow in real problems, these conditions can be barely prescribed because data $\boldsymbol{v}$ and $\boldsymbol{d}$ are often not available or measurable and they cannot be derived by physical arguments. In fact, in most of the cases we have insufficient data to make the problem well posed. In this case, we say that the boundary conditions are "defective". The numerical treatment of defective conditions is crucial and currently subject of active research, as the accurate selection of methodologies for their prescription is critical for the reliability of the numerical modelling. To be more concrete, we consider the example of the flow rate. The flow rate through a boundary section $\Gamma$ is technically defined as

$$Q(t) = \rho \int_\Gamma \boldsymbol{u}(\boldsymbol{x}, t) \cdot \boldsymbol{n} \, d\boldsymbol{x}. \tag{4}$$

This is an integral condition over an entire portion of the boundary $\Gamma$, available from measures, clearly "defective" as it is not providing pointwise data on each point of the boundary. Another practical case is the prescription of the pressure $p_m(t)$ retrieved from some measures and prescribed over $\Gamma$ in the form

$$p(\boldsymbol{x} \in \Gamma, t) = p_m(t) \tag{5}$$

Again, we have a scalar condition prescribed over the boundary in place of the three data required in each point $\boldsymbol{x} \in \Gamma$.

There are several practical ways for filling the gap in an engineering and mathematically consistent way. Here, we limit to a short summary of some possible approaches for the prescription of the flow rate. The reader interested in this topic is referred to [Quarteroni et al., 2016, Formaggia, Quarteroni et al. 2009] and the references quoted there.

As the flow rate $Q(t)$ is a defective condition, we need a way to fill the gap with the required data, that is a velocity profile $\boldsymbol{v}(\boldsymbol{x}, t)$. An engineering approach for this step is the introduction of an arbitrary, yet reasonable velocity profile $\boldsymbol{v}_a(\boldsymbol{x}, t)$ such that

$$\rho \int_\Gamma \boldsymbol{v}_a(\boldsymbol{x}, t) \cdot \boldsymbol{n} \, d\boldsymbol{x} = Q(t).$$

In most of the cases, this profile is the parabolic function of the Poiseuille-Hagen solution, or a Womersley profile when $\Gamma$ has a circular shape. In some circumstances, a flat velocity profile is a more reasonable option (e.g., at the entrance of the ascending aorta). The main advantage of this approach relies on its easiness, as it can be immediately applied with standard CFD solvers. The main drawback is the arbitrary selection of a profile that is impacting the solution in the region of interest. In fact, to mitigate the arbitrariness on the solution, an artificial elongation of the region of interest called *flow extension* is applied to the computational domain. This is intended to position the prescription of the arbitrary profile far from the region of interest so to ultimately dump the impact of the arbitrary choice on the numerical solution. This approach, partially justified by the theory developed in [Veneziani, Vergara, 2007] stating that the arbitrariness of the velocity profile exponentially decays within the volume of interest, is extremely popular even if the accurate construction of flow extensions is not necessarily a trivial step, in particular when working on a large volume of patients, like in Clinical Trials. Other approaches, mathematically more sound, are possible. In particular, we recall here two alternative methods.

1) The Lagrange multiplier approach [Formaggia et al., 2002, Veneziani, Vergara 2004, Veneziani, Vergara, 2007]. In this case, the flow rate is not regarded as a boundary condition, but as a constraint to prescribe to the Navier-Stokes equations. As such, a Lagrange multiplier approach can be pursued, working on the variational formulation of the problem. In this way, no velocity profile needs to be selected, as it results as a by-product of the Lagrange multiplier calculation. As a benchmark, when either a Poiseuille or Womersley profiles are the exact solution of the problem, they are correctly retrieved by this approach. The main drawback of this formulation consists of the additional cost required by the Lagrange multiplier. Approximate effective solution methods have been proposed [Veneziani, Vergara, 2007].

2) The Data Assimilation approach [Formaggia et al. 2008, Formaggia et al. 2010]. In this case, the condition on the flow rate is used for constructing a minimization procedure. The entire solution of the fluid dynamics is reformulated as the minimization of the mismatch between computations and data, under the constraint of the fluid equations.

The latter approach actually defines a change of the perspective about the considerations of boundary conditions. We do not add conditions to a set of equations, but "assimilate" available data to the mathematical model, in such a way that the results of the simulation match the available data [Asch et al., 2016, Law et al., 2015]. This requires the identification of variables to tune to attain the minimization, that we call "control variables". For instance, we can select the normal stress (or traction) $\boldsymbol{\tau} \equiv p\,\boldsymbol{n} - \nu\,(\nabla\,\boldsymbol{u} + \nabla^T\,\boldsymbol{u}) \cdot \boldsymbol{n}$ on $\Gamma$ as control variable and then solve the problem

*Find $\boldsymbol{\tau}$ such that*

$$\left( \rho \int_{\Gamma} \boldsymbol{u}(\boldsymbol{\tau}) \cdot \boldsymbol{n} - Q \right)^2 \; is\; minimal$$

*under the constraint of (1).*

This approach has been explored in [Formaggia et al. 2008, Formaggia et al. 2010], also in the case of fluid-structure interaction problems, with excellent numerical results. In fact, the approach is very general and applied to any kind of available data, regardless of their nature of incompleteness. In addition, it enforces the condition in a least square sense, which is particularly appropriate when measured data are noisy. The selection of the control variables is critical for the quality of the results obtained after the minimization, however, it is quite free and customized for the different problems at hand. The main drawback of this approach is the computational cost. The mathematical nature, in fact, attains to the class of inverse problems that typically require intense iterative processes. Approximate solution methods and reduced order modeling are mandatory in this case.

More in general, the prescription of boundary conditions is made complicated by the "multiscale" nature of the circulatory system. The conditions of the network downstream a region of interest may affect the hemodynamics also in that region. This is a physical feature of the circulation, which guarantees resilience to the functioning of the system in presence of occlusions or other pathologies. An occlusion is generally compensated by collateral pathways (that regularly may be listed as redundant) and, more in general, circulation features baroreflex and chemoreflex mechanisms that regulate the blood supply from the large vessels for the needs of the peripheral organs. The mathematical model should include, in this case, the presence of a peripheral circulation downstream a region of interest and this can be typically attained by coupling the 3D equations (1) with surrogate models. A popular strategy is the coupling of the Navier-Stokes equations with lumped parameter models [Quarteroni et al., 2016, Formaggia, Quarteroni et al. 2009], or, more in general, surrogate models. The numerical coupling of full and reduced models raises some methodological challenges investigated in the literature (see [Quarteroni et al., 2016] and the references there). In this section, to be concrete but with no claim of completeness, we address a popular simplified lumped parameter model that potentially captures the most relevant features of peripheral circulation, the so-called 3 Element-Windkessel model (3WK).

The name Windkessel comes from a German firemen tool to convert periodic into continuous flow, as it actually happens in the peripheral districts. In particular, the 3WK features three parameters, two representing viscous resistances and one the global effects of vessel compliance of the peripheral districts. By exploiting the usual analogy between hydraulic networks and electrical circuits, the model is represented by the scheme reported in Figure 5 where $R_1$ and $R_2$ represent the viscous terms and $C$ the compliance.

Figure 5. Diagram of the 3WK lumped parameter model for peripheral circulation.

With the notation of the figure, the underlying mathematical model reads

$$\dot{P_p} + \frac{P_p}{CR_2} = \frac{Q}{C}, P = R_1 Q + P_p.$$

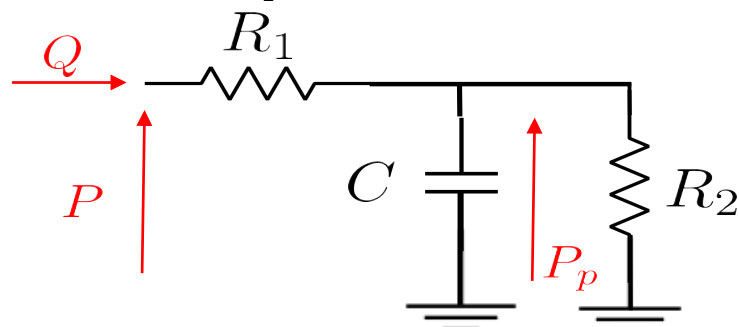

By a standard application of the method of integrating factors, and assuming that the solution is known at the instant $t_0$, we obtain

$$p_{3WK}(t) = p_{3WK}(t_0)e^{\frac{-t-t_0}{CR_2}} + R_1\left(Q(t) - Q(t_0)e^{\frac{-t-t_0}{CR_2}}\right) + \frac{1}{C}\int_{t_0}^t Q(\tau)e^{(\tau-t)/CR_2}d\tau.$$

This relation can be used as a condition to prescribe typically at outlet boundaries, to incorporate the presence of the downstream circulation in the form (recalling (3))

$$p\boldsymbol{n} - \nu(\nabla\boldsymbol{u} + \nabla^T\boldsymbol{u}) \cdot \boldsymbol{n} = p_{3WK}\boldsymbol{n}.$$

Recalling the definition of flow rate, this is, in fact, a condition combining pressure and velocity in a nontrivial way (*Robin condition*). In practice, one of the critical aspects when using this kind of models is the estimation of the parameters $R_1$, $R_2$ and $C$ in patient-specific settings. As it is currently possible retrieving realistic values of parameters from the literature, it has been recognized that the reliability of patient-specific simulations strongly rely on the appropriate quantification of those parameters based on the available data. In fact, this is still subject of active research and data assimilation techniques like Kalman-filtering or variational approaches are currently under investigation (see e.g. [Bertoglio et al. 2012, Bertoglio et al., 2014, Romarowski et al., 2017]).

### 3.2 A Survey of Numerical Methods for incompressible fluids

The incompressible Navier-Stokes Equations (1) still represent a mathematical challenge in many respects. Many fundamental theoretical problems are still open for this system. In practical settings, the solution of the problem given by (1), completed by initial and boundary conditions, needs to be obtained by numerical approximation, as analytical methods fail. Also, the numerical solution of these equations is challenging for the intrinsic features of the equations. In cardiovascular problems, the challenge is often burdened by the complexity of the geometries at hand. This subject is one of the pillars of computational fluid dynamics and we do not intend to be exhaustive here. We refer the reader to specific textbooks (see e.g. [Ferziger, Peric, 2012]). Here, we just recall the basic features of the numerical discretization of the problem.

The equations (1) are a nonlinear system of four partial differential equations depending on space and time. The backbone of the numerical discretization is to approximate these four equations with a (typically large) algebraic system, as many effective methods for solving algebraic systems of large dimensions with modern computers are available. This procedure ideally should be able of combining efficiency (the algebraic system is solved in a reasonable time) and accuracy (the approximation error is under control). Since we have two independent variables, space and time, the discretization is usually performed with respect to the two sets separately, as the differential nature of the problem is different in time and space.

*Space discretization* can be performed in many different ways. *Finite Differences* are the most intuitive method, based on a replacement of the derivatives with incremental quotients after having collocated the original problem on a suitable grid of points. These points are usually selected along the Cartesian directions and represent the world of the numerical solution, i.e., where the numerical solution is actually computed. The underlying tool for estimating the errors, in this case, is given by the Taylor expansion of the solution according to the different Cartesian directions. In spite of the simplicity, the method suffers from some drawbacks in particular for the lack of a rigorous complete mathematical background (Taylor expansion can be used only in presence of a regularity of the solution that is seldom verified in real problems) and for the intrinsic "Cartesian" footprint, that makes challenging the treatment of complex geometries.

Another approach extremely popular in this field is the *Finite Volume* method. For such a methodology, the volume of interest is split in many subdomains (or finite volumes) where the equations are integrated. By an extensive use of the Gauss theorem to reformulate volume to surface integrals and of quadrature approximations, the system is approximated by a system computing the average of the solution on each volume. The method features a significant simplicity and efficiency. In fact, it is the method of choice of many commercial packages. Nevertheless, the set-up of competitive high-order finite volume schemes may be tricky.

From the mathematical point of view, the schemes with the strongest theoretical background belong to the family of the *Galerkin methods* that includes *Finite Element* and *Spectral Methods*. In fact, these methods directly rely on the so-called *weak formulation* of the problem, stemming from the application of variational principles (virtual works). This is also the appropriate framework for theoretical analysis dealing with the possible lack of continuity in the solution - as it may occur in some applications. According to these methods, the numerical solution is not obtained by approximating the derivatives (or the integrals of the Finite Volume formulation) but by postulating a specific shape for the solution described by a finite number of parameters. For instance, in linear Finite Elements, the solution is assumed to be piecewise linear over a set of volumes or *elements* covering the region of interest. Likewise, one can consider piecewise quadratic approximations or more. As the derivative[1] of these functions can be promptly computed to obtain a system for the parameters describing the approximating solution. The approximation, in this case, is in the set (or, more precisely, the *functional space*) where the numerical solution is looked for. Galerkin methods can be analyzed with the tools of Functional Analysis and Approximation Theory, that led to a pretty complete picture of the different properties of Finite Elements in terms of accuracy and efficiency (https://femtable.org/). Similarly, spectral methods postulate the solution to be polynomial defined on special set of points (Gaussian nodes) that guarantee high accuracy. Unfortunately, the management of these nodes in real geometries may be troublesome. Spectral elements are a sort of compromise, merging the advantages of spectral methods in a decomposition of local subdomains or macro-elements. As a general statement, from the theoretical point of view, finer is the reticulation of the region of interest (the so-called mesh) and more accurate and numerically stable is the solution. In practice, under a certain level of refinement, the solution does not sensibly improve while computational costs and sensitivity to rounding errors increase. A good simulation practice is then to identify a mesh such that the solution does not show significant improvements to refinements (the so-called *mesh-independence* test). As an example of the fact that research on the development of general methods is still very active, we briefly mention here also *isogeometric analysis*, which has been already introduced and discussed in section 6.2.

---

[1] In fact, the concept of derivative that must be used here is a generalization of the classical definition, and is called generalized or distributional derivative (see, e.g., [Brezzi, Gilardi, 1987]).

For the *space discretization* of the Navier-Stokes equations, we focus here on using the Finite Element method, recalling some key aspects. An extensive study can be found, e.g., in [Elman et al., 2014] and the bibliography therein.

One of the fundamental features of the incompressible Navier-Stokes equations is the nature of *saddle point* problem. With this, we mean that the equations are the result of a minimization procedure constrained by the incompressibility. In this respect, *pressure is the Lagrange multiplier of the constraint*. This circumstance has a primary impact on the numerical discretization by finite elements. In fact, the correct formulation of the numerical approximation must obey some constraints on the selection of the trial space for the velocity and the pressure. This constraint goes generally under the name of *inf-sup* or *Babuška-Brezzi* condition [Brezzi, 1974]. From the practical point of view, this condition enforces the trial space of the velocity to be "large enough" with respect to the one for the pressure. For instance, a piecewise linear finite element approximation for both velocity and pressure is not viable, as the resulting algebraic system is singular. Piecewise quadratic velocities can be coupled to piecewise linear pressures to get a correct numerical approximation, however, this comes at the price of a large algebraic system to solve, with the consequent computational burden. A possible workaround is given by numerical stabilization techniques [Hughes et al. 1986], that, however, may affect the accuracy of the solution. The $P2 - P1$ pair (quadratic velocity, linear pressure), also called *Taylor-Hood element*, is, in fact, one of the most popular choices in biomedical applications.

The result of the space discretization is a Differential-Algebraic system that requires a time-discretization. The latter is usually worked out by finite difference schemes, for their simplicity and efficiency. Nevertheless, space-time finite element discretizations are possible [FSV]. With finite differences, the time derivative of velocity is replaced by incremental quotients, still formally based on Taylor expansions, and collocated at instants $t_k$ of the time interval, with step $\Delta t$. For simplicity, the step size is here assumed constant, yet it can be automatically adapted to guarantee accuracy during fast transients and efficiency when the dynamics is slow and less time samples are needed [Veneziani, Villa, 2013].

After time discretization, the system is eventually an algebraic system. Let $\boldsymbol{U}^k$ be the vector describing the velocity in the finite element space at time $t_k$ and $\boldsymbol{P}^k$ the pressure at the same instants. The algebraic system to be solved at each time step, for the time index $k$ running from 1 to a final step $K$, reads

$$\begin{bmatrix} A(\boldsymbol{U}^k) & B^T \\ B & 0 \end{bmatrix} \begin{bmatrix} \boldsymbol{U}^k \\ \boldsymbol{P}^k \end{bmatrix} = \begin{bmatrix} f(\boldsymbol{U}^{k-1}) \\ 0 \end{bmatrix}. \tag{6}$$

The matrix $A$ collects the discretization of the momentum equation, while $B$ is the discretization of the divergence operator. The function $f(\cdot)$ collects the action of source terms and boundary conditions. This is an algebraic nonlinear system and the final step to obtain the numerical solution is a linearization. This can be attained by iterative methods (like *Newton* or *Picard*), that however add another level of iterations in the time loop, or by extrapolation, i.e., by replacing $A(\boldsymbol{U}^k)$ with a matrix

$A_L = A(\boldsymbol{U}^*)$ where $\boldsymbol{U}^*$ is a time extrapolation of the velocity based on the previous solutions. The latter is usually preferred in computational hemodynamics for its simplicity and computational efficiency. After linearization we are left with a linear system in the form

$$\begin{bmatrix} A_L & B^T \\ B & 0 \end{bmatrix} \begin{bmatrix} \boldsymbol{U}^k \\ \boldsymbol{P}^k \end{bmatrix} = \begin{bmatrix} \boldsymbol{f}(\boldsymbol{U}^{k-1}) \\ \boldsymbol{0} \end{bmatrix}. \qquad (7)$$

This system is usually large – as required by the fulfillment of the inf-sup condition – and with spectral properties (i.e., the eigenvalue distribution) that make the efficient numerical solution quite challenging. In addition, it has to be solved at each time step, creating a global computational intensive problem. To tackle efficiently the solution, a first necessary requirement is the identification of an appropriate preconditioner, i.e. of an invertible matrix $P$ such that the system $P^{-1} \begin{bmatrix} A_L & B^T \\ B & 0 \end{bmatrix} \begin{bmatrix} \boldsymbol{U}^k \\ \boldsymbol{P}^k \end{bmatrix} = P^{-1} \begin{bmatrix} \boldsymbol{f}(\boldsymbol{U}^{k-1}) \\ 0 \end{bmatrix}$ presents much better numerical properties for the solution. Unfortunately, the choice of a good preconditioner, in real problems featuring a large size (order of millions or more), is generally not enough for an efficient solution. A possible workaround to reduce the cost is then to split the solution of velocity and pressure, so to break down the computational complexity into a sequence of smaller and easier problems. The splitting can be carried out following different lines, the literature on this topics is huge [Elman et al., 2014]. Here, we just recall two of them.

i) Chorin-Temam methods (split-then-discretize): Stemming from a well-known principle of differential vector calculus, called Hodge decomposition or Ladhyzenskhaja Theorem, the velocity solution can be regarded as the divergence-free component of a generic vector. The splitting is obtained by first computing this vector, and then by projecting it on the divergence-free space by subtracting the gradient of a scalar field related to the pressure. Overall, the solution is broken into a standard advection-diffusion-reaction problem and a Poisson problem for the pressure. Each problem can be discretized by standard finite element or other methods. The splitting is only approximated for the presence of the nonlinear convective term and boundary conditions required for the pressure (and not required by the original unsplit problem). An extensive theory on this method and its accuracy has been developed e.g. in [Prohl, 1997, Guermond et al., 2006].

ii) Algebraic splittings (discretize-then-split): In this case, the splitting is the result of an approximate factorization of the matrix of the system into triangular factors (I is the identity matrix)

$$\begin{bmatrix} A_L & B^T \\ B & 0 \end{bmatrix} \approx \begin{bmatrix} A_L & 0 \\ B & S \end{bmatrix} \begin{bmatrix} I & \tilde{B}^T \\ 0 & I \end{bmatrix}$$

where $S$ and $\tilde{B}^T$ have to be properly designed. In this way, the original system is naturally split into a sequence of systems alternatively for velocities and pressure [Perot, 1993, Veneziani, Villa, 2013]. The algebraic splitting approach has a less established theoretical background than the differential one, but it features comparable accuracy and it does not have to specify any boundary condition for the pressure, as they are automatically incorporated in the discretization step.

In the specific field of computational hemodynamics, popular strategies rely on splitting methods, either differential or algebraic, the latter being preferred for their versatility for the different types of boundary conditions we need to consider.

### *3.3 Modeling Turbulence*

The flow regime featured by blood is generally dominated by the pulsatility. High convective terms, measured by the presence of high values of the Reynolds number $Re = \rho\, UL/\mu$, where $U$ is a characteristic velocity of the problem and $L$ a characteristic length, may develop along the heartbeat, but typically do not last enough to trigger turbulent dynamics. Still, the presence of highly disturbed flow, as it may occur in the ascending aorta, generates numerical instabilities in the approximation process, that generally are sorted out with an accurate reticulation (Direct Numerical Simulation – DNS). However, in presence of some pathological conditions associated with locally high convective terms, there is an energy cascade from macro- to micro-scales in the fluid that require a specific modelling. Among the others, we mention Aortic Dissections, where the entry tears between the false and the true lumen determine a strong acceleration of the flow; the Total Cavopulmonary Connection – a special paediatric surgery that typically features a cross-shaped connection of vessels (e.g., [Mirabella et al., 2013]) - where the occurrence of colliding streams increases the relative Reynolds number between the incoming flow jets.

Turbulence may be described in different ways, according to the different applications associated with a different range of the Reynolds. When an appropriate refinement of the computational grid is not doable to embrace all the significant microscales involved in the dynamics, the backbone of numerical modeling relies on an appropriate surrogate modeling of the effects of microscales below the reticulation (unresolved) to the medium and large scales solved by the adopted mesh. Reynolds-Averaged Navier-Stokes models (RANS) are suitable for large Reynolds numbers and rely on time-averaged semi-empirical procedures. For moderate Reynolds numbers (up to 5,000), Large Eddy Simulations (LES) provide generally more accurate solutions. The surrogate modeling of the unresolved scales can be accomplished by solving an appropriate set of equations. For instance, in the Leray model, the original Navier-Stokes Equations are replaced by a more complex (and computationally demanding) system:

$$\begin{cases} \rho(\partial_t \boldsymbol{u} + (\boldsymbol{w}\cdot\nabla)\boldsymbol{u}) - \mu\nabla\cdot(\nabla\boldsymbol{u} + \nabla\boldsymbol{u}^{\mathrm{T}}) + \nabla p = \boldsymbol{f} \\ \nabla\cdot\boldsymbol{u} = 0 \\ \boldsymbol{w} - \delta^2\nabla\cdot(a(\boldsymbol{u})\nabla\boldsymbol{w}) + \nabla\lambda = \boldsymbol{u} \\ \nabla\cdot\boldsymbol{w} = 0 \end{cases}. \qquad (8)$$

where $\boldsymbol{w}$ and $\lambda$ are auxiliary functions filtering the original set of equations, $\delta^2$ is a parameter empirically selected (called *filter radius*) as a function of the reticulation size and $a(\cdot)$ is an empirically design function (called *indicator*). The rationale of this approach is that the indicator function $a(\boldsymbol{u})$ is $\approx 0$ for low Reynolds and $\approx 1$ for high Reynolds, so to activate a filtering action on the convective field $\boldsymbol{w}$ occurring in the LES equations. For low Reynolds, $\boldsymbol{w} \approx \boldsymbol{u}$, so to be consistent with the original Navier-Stokes equations (1). For high Reynolds, the indicator function stabilizes the numerical solver surrogating the effect of the high frequencies in the flow. Many different possible indicator functions are available, based on both mathematical and physical arguments [Bertagna et al., 2016]. Among the latter, we recall the Smagorinsky filter, while in the former group we recall the deconvolution low-pass filtering techniques. Since the computation of the indicator function may require the solution of an additional differential problem. Treatment of turbulence introduces then additional computational costs and specific methods for the efficient solution of these problems are currently subject of research.

## 4.0 Simulating and Reducing the Complexity

### 4.1 Fluid-Structure Interaction

A fluid-structure interaction problem in hemodynamics arises when we have a problem represented by the Navier-Stokes equations (1) coupled to a structural problem that generically we assume to be represented by, e.g., the equations of (nonlinear) elasticity. In general, the fluid and the solid domains are distinct and are indicated as $\Omega_f$ and $\Omega_s$, respectively. The two problems are coupled by interface conditions prescribing the continuity of the velocity of displacement in the solid and the velocity of the fluid and of the normal stresses insisting at the interface $\Gamma$,

$$\boldsymbol{u}_f = \boldsymbol{u}_s$$

$$\boldsymbol{\sigma}_f \cdot \boldsymbol{n}_f + \boldsymbol{\sigma}_s \cdot \boldsymbol{n}_s = 0 \quad \text{on } \Gamma, \qquad (9)$$

where $\boldsymbol{\sigma}_f$ and $\boldsymbol{\sigma}_s$ are the fluid and solid stress tensors respectively. When solving numerically these problems, several additional challenges rise beyond the solution of the individual components (fluid and structure). These challenges are diverse as different conditions of interaction between fluid and structure are faced in practice. In cardiovascular flows, we have two different main kinds of interactions that can be tackled with different approaches: (1) The interaction between the vascular wall and the blood flowing in; (2) The interaction between the blood and a body floating internally, like for the leaflets of valves at the entrance of the aorta, the mitral valve or the venous valves preventing backflows. The two problems present different features and are, typically, solved by different approaches.

Numerical methods in this field can be categorized in different ways. We first distinguish *monolithic* vs *partitioned* (or *segregated*) approaches. In the first case, the two problems are solved simultaneously, as a unique problem resulting in one linear system after all the needed procedures of discretization/linearization. This approach suffers from two drawbacks: (i) the coupled problem joins subproblems with different time constants and this reflects in bad conditioning properties of the numerical approximation; (ii) the number of degrees of freedom is generally huge, so the linear systems to solve at each time step are challenging and demanding for high-performance computing facilities. The two aspects call for appropriate preconditioners. In segregated approaches, a modular formulation is undertaken and the two problems are generally solved sequentially, by replacing interface conditions (9) with appropriate boundary conditions for each subproblem. The coupling of the two problems, then, can be categorized as *strong* or *weak* depending on the way the interface conditions are enforced at each time step. If (9) are fulfilled approximately (as function of the discretization parameters, for instance, the time step $\Delta t$), then we say that the coupling is *weak*, as it is exact only in the limit of the discrete problem tending to the continuous one. If in the discrete partitioned problems the enforcement of (9) is exact, we say that the coupling is *strong*. Another classification refers to the *reticulation* for the two subdomains. Assuming, for instance, to solve the two problems sequentially with the method of finite elements, we need to introduce a mesh for each problem. We say that the method is *fitted* if the degrees of freedom at the interface are shared by the two solvers. *Unfitted* methods are the ones where the degrees of freedom at the interface do not necessarily coincide and special procedures are needed to transfer information from one to the other.

To be more concrete, we exemplify the cases of weak and strong coupling by means of pseudo-codes (where we adopt names reflecting the notation used so far). We first present the pseudo-code sequence for the case of weak coupling.

```
ALGORITHM 1: WEAK COUPLING
Initialize: set u_f_onGamma
While (t<=tfinal){
    sigma_f = solveFluid(omega_f,u_f_onGamma)
    EnforceIC(sigma_s_normal=sigma_f_normal);
    u_s_onGamma = solveStructure(omega_s,sigma_s_normal);
    EnforceIC(u_f_onGamma=u_s_onGamma);
    ManageGeometry(omega_s,omega_f)
    t=t+Deltat
}//end while
```

In this sequence, we use the continuity of velocities as a boundary condition for the fluid. At the end of the fluid run, we retrieve information on the normal stress that becomes the surface force term on the boundary for the structure problem. Notice that, at each step, we need some geometry manipulation to manage the transfer between degrees of freedom of the two solvers (**EnforceIC**) and the change of configuration in the domains of interest (like a change of topology for a contact between two solids – **ManageGeometry**). Also, it is evident that the interface conditions are not prescribed exactly, as clearly in this implementation the continuity of the normal stress is, in fact, prescribing the fluid normal stress to equate the structure normal stress at the previous time step.

Instead, in the strong coupling case, we may have the following pseudo-code.

In this case, at each iteration, we sub-iterate the fluid and structure solvers to obtain the exact enforcement of (9). Clearly, this second approach is computationally more intensive, as at each step we need to iterate, each iteration requiring the solution of a fluid and a structure solver. The convergence of these iterations needs, in fact, to be proved with arguments of functional and numerical analysis (fixed point theorems). However, we will see that the weak coupling approach may suffer from instabilities that make it an unviable option.

Another categorization refers to the frame of reference from which the problem is written. Generally speaking, we may consider the two classical approaches: the Eulerian one, where the problems are written in a fixed frame of reference, and the Lagrangian frame of reference, where the frame of reference is evolving with the domain of interest, either $\Omega_f(t)$ or $\Omega_s(t)$. The two subdomains can be approached with different approaches, as they have both pros and cons. However, there are other possibilities that can be numerically more effective. This is exactly the case of the Arbitrary Lagrangian-Eulerian (ALE) methods that are popular for simulating the interaction between blood and vascular walls. In the Immersed Boundary methods,

```
ALGORITHM 2: STRONG COUPLING
Initialize: set u_f_onGamma
While (t<=tfinal){
 Converge=false;
  while (convergence==false){
      sigma_f = solveFluid(omega_f,u_f_onGamma)
      EnforceIC(sigma_s_normal=sigma_f_normal);
      u_s_onGamma = solveStructure(omega_s,sigma_s_normal);
      EnforceIC(u_f_onGamma=u_s_onGamma);
      ManageGeometry(omega_s,omega_f)
      convergence=TestConvergence(
                 abs(u_f_onGamma-u_s_onGamma),
                 abs(sigma_s_normal-sigma_f_normal));
  } // end while for the convergence
```

usually preferred for simulating floating bodies in fluids, a mixed Eulerian (fluid)/Lagrangian (structure) approach is instead preferred.

### *The Arbitrary Lagrangian-Eulerian Method*

The formulation of ALE methods dates back to the eighties [Hughes, Zimmerman, 1981; Donea, Giuliani, 1982]. The rationale of the method is that for the vascular wall it makes sense to approach the problem on a Lagrangian framework, being the deformations quite limited, but for the fluid problem, this may be impossible. In fact, in a Lagrange approach, the nodes of the finite element mesh move with the velocity of the fluid. In presence of flow recirculations (as it may happen downstream a stenosis), this rapidly impairs the quality of the mesh. On the other hand, if we treat the structure with a Lagrange frame of reference, it may be convenient to stick to the same reference at the interface between the two domains. For instance, if we want to use a fit-grid approach, we can move the interface nodes with the structure. The idea is therefore to introduce another frame of reference that is neither purely Lagrangian nor Eulerian: it moves with the structure at the interface, yet it is Eulerian at the other portions of the fluid boundary (like inflows and outflows) so to preserve the geometry of $\Omega_f(t)$ in time. This requires to: (i) introduce a new displacement field $\boldsymbol{\eta}_g$ of the grid and its velocity $\boldsymbol{w}_g = \dot{\boldsymbol{\eta}}_g$; (ii) rewrite the fluid problem with respect to the moving frame of reference featuring velocity $\boldsymbol{w}_g$. At this point, the displacement field must obey the following constraints

$$\boldsymbol{w}_g = \dot{\boldsymbol{\eta}}_s \text{ on } \Gamma, \qquad \boldsymbol{w}_g = \boldsymbol{0} \text{ on } \Gamma_{in} \cup \Gamma_{out}. \qquad (10)$$

Clearly, these constraints can be obtained by multiple choices for the velocity $\boldsymbol{w}_g$. The selection of $\boldsymbol{w}_g$ is therefore arbitrary and it can be obtained to obey criteria convenient for the numerical solution. For instance, it may be selected so to minimize the grid distortion to guarantee the quality of the finite element solution [Donea, Giuliani 1982]. A reasonable choice, trading off computational simplicity and all the requirements, is to solve a Laplace problem (harmonic extension)

$$-\Delta \boldsymbol{w}_g = 0 \text{ in } \Omega_f.$$

with (10) as boundary conditions. This approach does not necessarily guarantee sufficient quality of the mesh, depending on the flow regimes and possible more refined approaches (not discussed here) may be necessary.

To write the Navier-Stokes equations on the arbitrary frame of reference moving with velocity $\boldsymbol{w}_g$, we need to recall the Reynolds transport theorem, correcting the Lagrange derivative for the fluid

$$(\partial_t \boldsymbol{u})_{ALE} = (\partial_t \boldsymbol{u}) + (\boldsymbol{u} - \boldsymbol{w}_g) \cdot \nabla \boldsymbol{u},$$

So that the ALE fluid problem then reads

$$\begin{cases} \rho \big( \partial_t \boldsymbol{u} + ((\boldsymbol{u} - \boldsymbol{w}_g) \cdot \nabla) \boldsymbol{u} \big) - \mu \nabla \cdot (\nabla \boldsymbol{u} + \nabla \boldsymbol{u}^{\mathrm{T}}) + \nabla p = \boldsymbol{f} \\ \nabla \cdot \boldsymbol{u} = 0 \end{cases}. \tag{11}$$

The numerical analysis of the solution of this coupled problem is clearly non-trivial. We refer to the literature for a comprehensive understanding of the topic [Nobile, Formaggia, 1999, Formaggia, Nobile, 2004, Gerbeau, Fernandez, 2009]). Here, we limit to recall some key items to help the reader to understanding the main problems of ALE methods.

*The Added-Mass effect*

When working in a weakly-coupling setting, numerical instabilities occur when the mass of the wall is significantly less than the mass of the fluid (as it happens in cardiovascular applications), the instabilities being more evident for long pipes. The reason of this effect was elucidated in the seminal paper [Causin et al. 2005], working on a simplified problem. The analysis gives a theoretical foundation to the instabilities, in terms of the eigenvalue analysis of an operator lumping in the structure problem the added mass impact of the fluid. This analysis goes therefore under the name of "added-mass effect" [Gerbeau, Fernandez 2009]. The analysis shows that the instabilities of the weak coupling can make it not an option for vascular problems. On the other hand, the strong coupling can be arranged to converge with appropriate numerical methods, yet with high computational costs as each. A trade-off is given by semi-implicit couplings, where, in fact, only part of the coupling is performed by sub-iterations, and the geometry management is done explicitly.

*Different treatment of the interface conditions and other splittings*

In the previous examples of segregated procedures, we simply associated one of the interface conditions (9) to the fluid problem and the other to the structural one, similarly to what is done in the Dirichlet-Neumann approach for Domain Decomposition methods. The iterative procedure reads each interface condition as a boundary condition for the associated problem. In fact, the (9) can be replaced by their linear combination with arbitrary parameters, so to provide a new set of interface conditions to enforce. While the new conditions are equivalent for the continuous problem, they are different in terms of numerical efficiency – as it is in Robin-Robin domain decomposition techniques [Quarteroni, Valli, 1999]. This idea was explored in [Badia et al., 2008] and the selection of the coefficients of the linear combination of the interface conditions that optimize numerical performances is investigated in [Gerardo Giorda et al., 2010].

Other, different ways for splitting the fluid and structural problem, by-passing the added-mass effect, were introduced in [Guidoboni et al., 2009].

*The Geometric Conservation Laws (GCL)*

In assessing the accuracy of moving domain problems, it is not possible to separate the accuracy analysis of space and time discretization, as the space domain is evolving in time and the way the discretization is performed has, in general, an impact on the space accuracy, too. For instance, the numerical differentiation to obtain the fluid velocity $\boldsymbol{u}_f$ from the boundary structural displacement $\boldsymbol{\eta}_s$ impacts the entire accuracy. To have a clear assessment of the interplay between time and space accuracy, the concept of Geometrical Conservation Laws (GCL) was introduced [Lesoinne, Farath, 1996]. These laws provides a set of rules to guarantee a certain order of accuracy to the solution as a function of the order of accuracy of the space discretization and of the numerical approximation of the grid motion in time (see for instance [Nobile, Formaggia 1999], Proposition 3.1).

### Immersed Approaches

#### Immersed Boundaries: Purpose and concepts

In many applications, the geometric complexity of the cardiovascular system is such that it is difficult to rely on traditional meshing strategies. With traditional meshing we mean here methods for which the entire geometry is built *a-priori* and/or for which the same type of approach is used to adapt the mesh (i.e., by remeshing the entire mesh). For instance, constrained Delaunay tessellation methods (see, e.g., [Frey, George, 2008, Cheng et al., 2012]) fall into this category as well as implicit function meshing (see, e.g., [Dapogny et al., 2015, Persson, 2004]). In this sense, this fits the classical pipeline for numerical simulations: pre-processing and then processing, where pre-processing contains the mesh generation part.

Immersed approaches are different since the geometrical features of the problem are handled within the processing part. Conceptually speaking, given the problem domain $\widetilde{\Omega} \subset \Omega$, the main idea consists of solving accurately the problem in $\Omega$. For this reason, most immersed approaches are closely related to interface problems since the first issues to arise are in the vicinity of $\partial\Omega$. Many immersed approaches relate to finite volume and finite difference methods, but here we restrict ourselves to basic Galerkin methods. In fact, there exist numerous methods which could be associated with the term "immersed approaches" and we will not provide an exhaustive discussion of all these methods. Rather the interested reader can delve into the references within the provided bibliography.

#### Historical note on immersed methods

Immersed approaches have a long history and could be traced back to [Hyman, 1952] (see, e.g., [Glowinski, 2003]). To the best of our knowledge, with respect to the finite element method, a mathematical analysis of such a type of method could be traced back to a study of Babuška [Babuška, 1970] on the finite element method with discontinuous coefficients. The term Immersed Boundary was first coined by C. Peskin in [Peskin, 1977] with application to blood flow in the heart, and the idea has been further developed in the 1980s until now [Peskin, 2002]. The Fictitious Domain method is instead to be traced back to the work of R. Glowinski et al. [Glowinski et al., 1994].

A typical Galerkin type boundary-value problem read:

Find $\widetilde{\mathbf{u}}_h \in V_h(\widetilde{\Omega}) \subset V(\widetilde{\Omega})$ such that

1. $a(\widetilde{\mathbf{u}}_h, \tilde{\mathbf{v}}_h) = b(\tilde{\mathbf{v}}_h)$    for all $\tilde{\mathbf{v}}_h \in V_h(\widetilde{\Omega})$, and

2. $\widetilde{\mathbf{u}}_h = \mathbf{g}$ on $\partial\Omega$.

One of the main issues here is how to enforce the second equation. We consider that there are two main categories of immersed approaches: 1) via basis modifications that is to modify $V_h(\widetilde{\Omega})$ such that it includes some knowledge of $\mathbf{g}$ and $\partial\Omega$. Methods that fall into this category are the Partition of Unity Method (PUM) and the eXtended Finite Element Method (XFEM), or Local Remeshing (LR); and 2) via a variational form modification, i.e., by modifying $a(\widetilde{\mathbf{u}}_h, \widetilde{\mathbf{v}}_h)$ or $b(\widetilde{\mathbf{v}}_h)$ such that they include some knowledge of $\mathbf{g}$ and $\partial\Omega$. Methods that fall into this category are, for example, the Finite Element Immersed Boundary Method (FE-IBM) or the Fictitious Domain Method (FDM). We will provide in subsequent sections a detailed list of references for each discussed method.

Notice that in most applications, a solution $\widetilde{\mathbf{u}}$ in $\widetilde{\Omega}$ will exhibit a singularity near $\partial\Omega$. Retaining an accurate method is challenging, and is still an active area of research.

*Functional space-based approaches*
The very idea of this class of methods is to explicitly provide in the finite element space information of the immersed boundary. Such operation can be either to add new basis functions, which are, for example, discontinuous and/or interpolatory on $\partial\Omega$. Local remeshing can be seen as a way to locally modify the discrete finite element spaces to accommodate for $\mathbf{g}$ and $\partial\Omega$.

*Basis modification*
Famous methods within this category are the Partition of Unity Method (see, e.g., [Melenk, Babuška, 1996]) and the eXtended Finite Element Method (see, e.g., [Sukumar et al., 2001]). For fluid-structure interactions problems, an XFEM method has been described in [Gerstenberger, Wall, 2008], and in this way, considering an aortic valve type problem, the velocity field would be continuous but the pressure discontinuous across the valve. We could consider a method [Buscaglia, Ruas, 2013] in which the pressure would be discontinuous across the valve. In this case, on the contrary of [Gerstenberger, Wall, 2008], the velocity gradient would be continuous, so we would reduce accuracy, but the method would be slightly less costly.

Other methods that fall in this category are the immersed interface method (see [Li, 1998, Lew, Buscaglia, 2008, Bastian, Engwer, 2009], to cite a few). Even if these methods have a good approximation of the singularity within $\widetilde{u}$, enforcing essential boundary conditions on $\partial\Omega$ is not straightforward and most of the time weak approaches are used, such as using Lagrange multipliers, Nitsche, or penalty approaches (see, e.g., [Hansbo, Hansbo, 2002, Hansbo et al., 2008] and references therein).

*Local remeshing*
Another "simple" approach is to consider a local remeshing. Advantages are that the fields are easily interpolatory at the interface and few degrees of freedom are added. Considering FSI problems such type of method has been used in [van Loon et al., 2006, Ilinca, Hetu, 2012, Auricchio et al., 2015, Frei, Richter, 2014].

In [van Loon et al., 2006] the mesh is modified locally but enforcing that the elements in its vicinity remains regular while in [Auricchio et al., 2015, Frei, Richter, 2014] anisotropic elements are employed. Such a type of elements introduces two issues related to conditioning and to the satisfaction of an inf-sup condition. Conditioning aspects are dealt in [Frei, Richter, 2014] for 2D problems, but a 3D extension is non-trivial. In [Auricchio et al., 2015] it was showed in 2D that inf-sup issues may arise with discontinuous pressure mixed elements, but they are very unlikely for most practical applications using continuous pressure elements. The interested reader can find more details in [Auricchio et al., 2016]. Conditioning issues also affect the XFEM method (see, e.g., [Zunino, 2011]).

Finally, we discuss the Finite Cell Method (FCM) [Parvizian et al., 2007]. The FCM is in many respects very similar to the PUM/XFEM method (in particular if one sees the XFEM as in [Hansbo2002]); the interested reader may also refer to [Joulaian et al., 2013]), extending its concepts to the *h-p* framework. An extension of the FCM [Rank et al., 2012] has also been proposed within the isogeometric framework [Cottrell et al., 2009].

*Variational formulation-based approaches*

On the contrary of basis modification-based methods, in this category of methods the discrete functional space is left untouched, which has the main advantage to maintain the same number of degrees of freedom of the problem. In general, a cost to such an approach is accuracy as the basis does not capture the singularities near the interface. Of course, such a statement should be taken with a grain of salt as the method presented previously are not (see [Boffi et al., 2008]).

We first present the Immersed Boundary Method and then the Fictitious Boundary Method.

The very idea of the Immersed Boundary Method (see, e.g., [Peskin, 2002]) is that "immersed" boundaries (or the action of the solid on the fluid) act as an external load, more precisely as a set of Dirac delta functions. Initially, the method was proposed using Finite Difference schemes. A variational formulation was proposed in, e.g., [Boffi, Gastaldi, 2003] and [Heltai, Costanzo, 2012]. Setting the problem within a variational formulation greatly simplifies the formulation since Dirac deltas are naturally handled within this framework.

The Fictitious Domain Method by Glowinski and coworkers (see, e.g., [Glowinski, 2003]) was initially set in the variational setting. In this method, the "immersed" boundaries (or the action of the solid on the fluid) are set using Lagrange multipliers (either boundary or distributed Lagrange multipliers can be used for thin or thick structures, respectively).

It is possible to reframe the Immersed Boundary Method as a Fictitious Domain Method as shown in [Auricchio et al., 2015, Boffi et al., 2015]. The interested reader may find more information and references in [Auricchio et al., 2014], also on the link between immersed approaches and elliptic interface problems.

Finally, the Immersogeometric method [Kamensky et al., 2015] is similar to the Fictitious Domain Method with stabilized Lagrange multipliers (via an augmented Lagrangian formulation) and, in this, it resembles Nitsche's method after elimination of the Lagrange multipliers. The main contribution of the work is to frame the Fictitious Domain Method within the isogeometric method [Cottrell et al., 2009].

## 4.2 Reduced Order Methods: an overview

Advanced applications in physics or engineering need an efficient solution of parametrized partial differential equations (PDEs), which requires the computation of the solution of (possibly nonlinear) PDEs for several different "scenarios". A solution by traditional methods like finite elements or finite volumes may not always be feasible. Reduced-order models have thus been studied to attempt the delivery of an accurate solution at lower computational costs [Quarteroni, Rozza, 2013]. When the problem is modeled with nonlinear equations several complexities arise to be faced in order to guarantee efficiency, accuracy, and reliability also in reduced-order modeling. Among them we mention the efficient exploration of the parameters space to build efficient reduced basis (low dimensional approximation spaces, but also representative ones to capture fine physical features and details (see e.g. [Constantine, 2015]), approximation stability to be guaranteed, arising, for example, in noncoercive problems, and especially to approximate vectorial multi-fields (i.e. pressure, velocity, temperature), accurate and fast estimation of stability factors (coercivity and inf-sup constants) as resumed recently in [Lassila et al., 2013a]. In cardiovascular problems, a combination of different ingredients characterizing reduced basis approach (such as online Galerkin projection, approximation stability by supremizers to enrich the approximation spaces) with POD (Proper Orthogonal Decomposition) features, which allows to build accurate reduced spaces without greedy procedures and error estimators, can be used with promising results. These combinations have been tested on nonlinear steady viscous flows modeled by Navier-Stokes equations and characterized by physical and geometrical (domain) parametrization. This strategy allows to offer a valid alternative approach preserving offline-online computational decomposition procedures, approximation stability and all the properties inherited by POD (such as orthonormalized basis functions for a robust algebraic stability and hierarchical spaces) as well as by the ones brought in dotation by Galerkin projection (orthogonality, best fit approximation, square effects) and the further possibility to develop an a-posteriori error analysis based on residuals and stability factors approximation to endow also POD with online error bounds. The reader can refer to [Lassila et al., 2013c] for all the methodology details. A brief historical excursus about methodologies, aspects and previous accomplished milestones taken into consideration in the development of this work follows.

POD was born to provide efficient model order reduction in turbulent viscous flows preserving the most important energetic features based on modal analysis and singular value decomposition (see, e.g., [Holmes et al., 1998, Aubry et al., 1988, Aubry, 1991, Cazemier et al., 1998, Berkooz et al., 1993, Ravindran, 2000]) with several further improvements addressed in the subsequent studies and developments (eigenvalues and eigenvector calculations, error estimation, physical parametrization beside time, optimal sampling, etc.), see for example several extensions proposed in (see, e.g., [Weller et al., 2010, Bergmann et al., 2009, Bergmann, Iollo, 2008, Iollo et al., 2000, Hay et al., 2009, Utturkar et al. 2005, Christensen et al., 1999, Kunish, Volkwein, 2003, Wang et al., 2012]). We also mention some very recent works [Caiazzo et al., 2014, Bergmann et al., 2013] for a state-of-the-art-review and references recalled in their bibliographies, among them we mention Window POD techniques, for example, to specialize the reduced basis and accomplish a certain efficiency in modeling more and more complex and extended systems [Grinberg et al., 2013]. In recent years a certain emphasis and growing attention has also been dedicated to stabilization techniques for POD [Iollo et al., 2000, Caiazzo et al., 2014, Akhtar, 2009], Sirisup, Karniadakis, 2005] and in the combination of Galerkin strategies with POD [Baiges et al., 2013, Chapelle et al., 2012, Chapelle et al., 2013, Amsallem, Farhat, 2013], to cite few recent contributions.

Going back now more specifically on Reduced Basis (RB) methods, they were proposed for nonlinear viscous flows since Eighties [Peterson, 1989] and a milestone contribution was provided by Anthony Patera and his group at MIT and collaborators at Paris VI (Yvon Maday) and in more recent years [Veroy, Patera, 2005, Nguyen et al., 2005, Quarteroni, Rozza, 2007, Deparis, Rozza, 2009, Deparis, 2008, Manzoni, 2014] with recent several applications [Lassila et al. 2013b, Manzoni et al., 2012a]. Starting from the linear saddle-point problem representing the Stokes system several studies were carried out to guarantee the stabilization of the reduced parametrized solution both for the global approximation with pressure recovery and algebraic aspects (extension of these aspects to Navier-Stokes nonlinear flows solved by reduced basis method has been straightforward, as in [Rozza, 2006]). The former aspect (approximation stability) was taken care thanks to supremizers enrichment to guarantee the existence of an equivalent parametrized inf-sup constant and the fulfillment of a related equivalent Brezzi condition. It was clear since first numerical tests that a combination of stable global approximated solutions was not always giving a stable solution, especially with geometrical parametrization (also in the linear case). We recall several options for this stabilization in order to propose efficient, stable and accurate methodologies [Rozza, Veroy, 2007, p. 200, Rozza et al., 2013, Gerner, Veroy, 2012] and to keep a stable and accurate approximation for the scalar pressure too. A certain importance is given to flows in parametrized domains for which the supremizers enrichment is crucial [Manzoni et al., 2012b]. This approach is different with respect to the one proposed previously in POD approaches with stabilization during the basis functions calculation (offline) and during the (online) ROM (for example using SUPG) [Caiazzo et al., 2014]. The latter aspect (algebraic stability) was focused on orthonormalization procedures of *Gram-Schmidt* [Rozza, Veroy, 2007]. More recently a third stability aspect, based on supremizers, has been introduced and studied for the enrichment of primal, dual and control reduced spaces for optimal flow control problems in a saddle point formulation [Negri et al., 2015].

RB framework for the approximation stability of flows in a POD setting is a very useful tool in order to improve the performance of the latter, already attractive and versatile, especially in parametrized domains. In this way, POD could benefit from a previously developed robust framework for the stabilization of viscous flows with computed pressure. At the same time POD could provide an alternative option to greedy algorithm in systems where the error bounds are not yet available or POD facilities are more promptly available (for example in already developed libraries/codes). The goal is to have several options and possibilities to combine reduced-order ingredients to be able to address important computational needs with the best possible compromise in terms of accuracy, costs, performances and fulfillment of additional need, e.g., error bounds.

A combination within POD and RB tools is not new, of course, for example, we mention the POD-greedy synergy approaches for unsteady problems [Haasdonk, Ohlberger, 2008, Nguyen et al., 2009] where the time evolution is captured by POD and physical and/or geometrical parameters are managed by greedy techniques.

Recent works in cardiovascular modeling and simulations with reduced order methods are devoted to fluid-structure interactions (see [Ballarin, 2015, Lassila et al., 2012, [Ballarin, Rozza, 2016, Ballarin et al., 2016, Bertagna, Veneziani, 2014, Wang et al., 2012]]), stability of flows and bifurcations (see [Pitton et al., 2017, Pitton, Rozza, 2017] as a references), optimization and control (e.g., see [Manzoni et al.,

2012c, Lassila et al., 2013a, Lassila et al., 2013b]).

Last, but not least, we mention two other reduction techniques: the Proper Generalized Decomposition (PGD, see [Chinesta et al., 2016] as a reference) and the Hierarchical Model Reduction (HIMOD) for flows (see [Baroli et al., 2016]). A recent collection with up-to-date research in reduced order method is [Benner et al., 2017].

### 4.3 Reduced Order Methods: general formulation

This section aims at introducing the reduced basis (RB) approximation for parametrized partial differential equations (PDEs), focusing on the Proper Orthogonal Decomposition-Galerkin (POD-Galerkin) approach. The interest in an efficient resolution method for parametrized PDEs arises in very different contexts (i.e. see [Boyaval et al., 2009, Dedè, 2010, Milani et al., 2008, Lassila et al., 2014, Rozza et al. 2013, Rozza et al., 2008]), and, specifically, it could be very useful in cardiovascular applications, from shape optimization and control problems (e.g., see [Manzoni et al., 2012c, Lassila et al., 2013a, Lassila et al., 2013b]), to fast blood simulations (e.g., see [Ballarin et al., 2016a,Ballarin et al., 2016, Manzoni et al., 2012a]) and fluid-structure interactions in hemodynamics (see [Lassila, 2012d, Ballarin, Rozza, 2016,Ballarin et al., 2016b] as references). This field of application is characterized by models governed by parametrized PDEs, possibly nonlinear. Computationally, cardiovascular models in hemodynamics could be very demanding: on one hand because of the PDEs governing blood flow (nonlinear Navier-Stokes equations, which need stabilization techniques in order to give reliable results), on the other one for the parametrization setting, which could be both physical and geometrical. To face those issues reduced order methods could be very attractive and versatile (an application of a stabilized reduced order method to Navier-Stokes equations is discussed in [Ballarin et al., 2015]). While only classical numerical methods, as finite element, could not lead to efficient results, a reduced order approach allows solving many-query cardiovascular problems in an accurate and reliable way. In this section, we propose a POD-Galerkin approach as a strategy to promptly handle these systems. This method is based on a POD procedure followed by a Galerkin projection onto the reduced space obtained. It has been already exploited with promising results in cardiovascular applications (see, for example, [Ballarin et al., 2016a, Ballarin et al., 2017, Ballarin, Rozza, 2016b])

In the following, the general idea of reduced order techniques and then POD-Galerkin method are introduced, in order to show a versatile high-performing numerical approximation for this specific field of application.

### Overview

In order to understand how reduced order methods work (see [Hesthaven et al., 2015],

[Quarteroni et al., 2015]  as references), let us introduce the *solution manifold* $\mathbb{M}$, defined as

$M=\{u(\mu): \mu \in \mathbb{P}\}$.

In other words, the solution manifold is the set of the solutions $u(\mu)$ of the parametrized PDE under the variation of the parameter $\mu$ in the space of the parameters $\mathbb{P}$. As in the continuous case, we can define the *approximated solution*

*manifold*, which is the set of the *full order* (i.e. finite elements) solution manifold under the variation of **μ**, that is

$$\mathcal{M}_h = \{u_h(\mu) : \mu \in \mathbb{P}\}.$$

The reduced order methods aim at building a reduced solution space exploiting specific values of the parameter μ, say *N* values of the parameter, in order to describe the *approximated solution manifold* in a reliable and fast way. Then, the reduced system is solved in a lower dimensional framework with respect to the *full order* one.

Let us indicate with *H* the dimension of the *full order* space. The RB approximation is based on two different stages:

1. **offline stage**: it is a (potentially) costly phase, where the solution manifold is explored in order to build a reduced space capable to describe with a sufficient accuracy any particular solution of $\mathcal{M}$. Computationally one has to solve *N* problems with *H* degrees of freedom;

2. **online stage**: it consists of a Galerkin projection onto the reduced basis space, for a particular parameter value. The computational cost of this phase is independent from *N.*

The online phase is performed every time we want to simulate our model for a new value of the parameter μ. The RB approach is efficient when the online stage is fast to perform: this is possibly true when *N* is much smaller than *H.* Being more specific, reduced order methods are applied in order to build reduced spaces, consisting in the linear combinations of the *full order* solutions evaluated in properly chosen *N* values of the parameter μ, in other words

$$V_N = \text{span}\{u_h(\mu_n) : n = 1, \dots, N\}.$$

There are essentially two classical approaches in order to build the reduced spaces: one is the so-called *greedy algorithm* (see section 3.2.2 of  [Hesthaven et al., 2015]), the other one is the *Proper Orthogonal Decomposition* (POD).

The greedy strategy is an iterative procedure which adds a basis function at each step in order to improve the reduced approximation. It requires the resolution of *N full order* problems. In order to apply a greedy algorithm, an a-posteriori error bound based on residuals must be available. At each iteration of the process, the parameter of evaluation of the true solution is chosen as the one that maximize this error bound (for all the technical details see [Boyaval et al., 2009] as a reference).

In the following we will focus our analysis on the latter of the two algorithms, since POD is an attractive prompt strategy to follow as sampling procedure:  while the greedy algorithm is based on the computation of error bounds which are still not always available for some systems, error bounds are not needed to perform a POD reduction.

**Problem Formulation**

In this section we will introduce the concept of parametrized PDE, focusing on the elliptic case.  Let P(μ) be our parametrized problem, with μ physical and/or geometrical parameter in the parameter space $\mathbb{P}$.  Let Ω be a two/three-dimensional physical domain and *V* a suitable Hilbert space endowed with the norm

$$(v,v) = \big((v)\big)^2.$$

Let us consider the parametrized *V*-continuous functionals

$$f(\mu) : V \to \mathbb{R},$$

and the  parametrized *V*-bilinear form

$$a(\mu): V \times V \rightarrow \mathbb{R}$$

The parametrized weak formulation of the problem reads: given μ in ℙ, find $u(\mu)$ in $V$ such that:

$$a(u(\mu),v;\mu)=f(v;\mu),$$

for all *v* in V. Under the assumption of the coercivity of $a(\mu)$, Lax-Milgram theorem guarantees existence and uniqueness of the solution $u(\mu)$.

Truth Approximation
This section aims at describing the abstract formulation of the discretized version of the parametrized problem ( P($\mu$) ). Let us consider the *H*-dimensional space $V_h \subset V$. The discrete version of the problem ( P($\mu$) ) reads: given μ in ℙ, find $u_h(\mu)$ in $V_h$ such that

$$a(u_h(\mu),v;\mu)=f(v;\mu),$$

for all *v* in $V_h$. If the dimension *H* of the discretized space is quite high, then the problem could be very costly, computationally speaking. If Lax-Milgram theorem is verified for the continuous version of (P($\mu$)), then it holds also for the discretized version and this ensure existence and uniqueness of $u_h(\mu)$. In the following we, will refer to the discretized solution with *full order solution* or *truth solution*. The approach described is the so-called Galerkin-projection method onto the discretized space $V_h$.

**Affine Decomposition**
To ensure the efficiency of the reduced order methods, the so-called *affine decomposition* has to be verified. This assumption is essential to guarantee an adequate Offline-Online procedure, which will be described in the following. We are assuming that the bilinear form
 $a(\mu)$ and the linear form $f(\mu)$ are affine in the parameter μ, that is: there exist $Q_a$ and $Q_f$ such that, for all the values of μ, the forms can be rewritten in the following way, respectively:

$a(\text{w,v}: \mu) = \sum_{q=1}^{Q_a} \theta_a^q (\mu) a_q(\text{w,v}), \qquad \forall \text{w,v} \in V$

 and

$f(\text{v};\mu) = \sum_{q=1}^{Q_f} \theta_f^q (\mu) f^q(v), \qquad \forall v \in V.$

In other words, we are assuming that the forms of the elliptic problem considered, could be written as the finite sum of μ-dependent scalar quantities and μ-independent forms. This assumption allows divide the reduction strategy in two stages, *offline* and *online,* which will be described properly in the next section.
In some physical systems the affinity assumption is not fulfilled and, then, one has to obviate through some numerical strategies and techniques. We mention the Empirical Interpolation Method (EIM, see [Barrault et al., 2004] as a reference), which allow to recovery the affine structure of the problem in order to guarantee the efficiency of the reduced order methods.

**Reduced Order Approximation and the Offline-Online Procedure**
In the introduction to this section, we have already presented fundamental concepts that are behind the reduced order methods (ROMs). Reduced strategies aim at approximating the *full order solution* $u_h(\mu) \in V_h$ in a reliable and fast way, exploiting a low dimensional framework. Reduced basis approach is based on a Galerkin projection onto a reduced space defined as:

$V_N$=span$\{u_h(\mu_n)$:n=1, ... ,N$\}$.

In other words, the method looks for a solution in a space made up by linear combinations of *truth solutions* evaluated in properly chosen parameters: in the next section we will present how this reduced space could be built, now let us assume that we have already built the reduced space verifying

$V_N \subset V_h \subset V$.

Exploiting this new approximated space, let us introduce the reduced problem in a Galerkin projection formulation onto the RB spaces. It is a new approximated problem and it reads: given $\mu$ in $\mathbb{P}$, find the reduced solution $u_N(\mu)$ in $V_N$ such that

$a(u_N(\mu)$,v;$\mu$)=f(v;$\mu$),

for all $v$ in $V_N$. We will refer to this reduced formulation as *reduced* problem.

Applying an orthonormalization technique in the inner product defined by *V,* the reduced space can be defined as:

$V_N$=span$\{\zeta_1, ... ,\zeta_N\}$.

In this new framework, the reduced solution in terms of the new orthonormalized reduced basis is defined as follows:

$$u_N(\mu) = \sum_{j=1}^{N} u_N^j(\mu)\zeta_j.$$

Substituting the latter expression in the *reduced problem* and choosing as test functions the basis functions the following algebraic reduced system holds:

$\sum_{j=1}^{N} a\left(\zeta_j,\zeta_i;\mu\right)u_N^j(\mu)$=f($\zeta_i$;$\mu$),     i=1,...,N.

Usually, the system is low dimensional, but its formulation is linked to the FE approximation space in the basis functions. If one assembles the reduced basis forms for every value of the parameter, the evaluation process will remain very expensive in terms of computational cost. Thanks to the affinity assumption, the assembling process can be decoupled in two phases: the Offline and the Online stages, that allows to efficiently solve the system for each new value of the parameter. Specifically, the system is rewritten as:

$\sum_{j=1}^{N}\sum_{q=1}^{Q_a} \theta_a^q(\mu)a_q(\zeta_j,\zeta_i)u_N^j(\mu) = \sum_{q=1}^{Q_f} \theta_f^q(\mu)f^q(\zeta_i)$, i=1,...,N.

Under this assumption is clear that a reduced basis approximation could be efficiently performed in two different stages: a $\mu$-independent costly Offline phase and a real-time $\mu$-dependent Online phase. The first procedure is needed only once, while the second process is applied at every new evaluation of $\mu$.

- In the **Offline stage,** the reduced space is built and orthonormalized. After this preliminary phase, all the $\mu$-independent quantities are assembled and stored.

- In the **Online stage**, the structures and forms evaluated and stored in the previous step are exploited in order to build the reduced system and, thanks to a Galerkin projection onto the reduced space, the reduced parametrized solution is computed.

This division of the assembling process allows the user to obtain very fast results and to simulate many times in a low dimensional and computational time/space saving framework.

**Proper Orthogonal Decomposition**

To apply a POD, a discrete and finite-dimensional subset of $\mathbb{P}$ is needed. For this specific set of parameter, one can define the *discrete approximated manifold*

$$\mathcal{M}_h(\mathbb{P}_h) = \{u_h(\mu): \mu \in \mathbb{P}_h\},$$

of cardinality *M*. When the discretization of $\mathbb{P}$ is fine enough, the *discrete approximated solution manifold* describes properly the approximated solution manifold, in other words:

$\mathcal{M}_h(\mathbb{P}_h)\approx\mathsf{M}_h$.

The POD algorithm is based on two processes:

1. sampling the discrete parameter space in order to compute the truth solutions at the chosen parameters,
2. a compression phase, where one discards the redundant information.

Under an algebraic point of view, the *N*-space resulting from the POD algorithm is built through the resolution of an eigenvalue-eigenvector problem on the correlation matrix of the *full order* solutions evaluated in *M* specific parameters. Let us consider the set of the solutions evaluated in these specific values of the parameters $\{u_h(\mu_m):\mathsf{m=1, ... ,M}\}$.

Then, let us define the real-valued the correlation matrix as

$$C_{\mathrm{mq}} = \frac{1}{M}\left(u_h(\mu_m),\mathsf{u}_h(\mu_q)\right)_V$$

where *V* is the solution space and $1 \leq m,q \leq M$. Then, the POD approach aims at solving the *N*-largest eigenvalue-eigenvector problem:

$$\mathsf{C}v_n=\lambda_n v_n$$

with $1 \leq n \leq N$. The eigenvector has been taken with unitary norm. Giving a descending order to the eigenvalues, the reduced space is build as

$$V_N=\mathrm{span}\{\xi_{1,} ... ,\xi_N\},$$

where the orthogonal basis is given by:

$$\xi_n = \frac{1}{\sqrt{M}} \sum_{\mathrm{m=1}}^{M} (v_n)_m \, u_h(\mu_m),$$

with $1 \leq n \leq N$.

This reduced space construction is performed during the offline stage. The online stage consists in solving the problem through a Galerkin projection onto the reduced space, obtaining a reliable (and fast) reduced solution.

**5.0 Conclusions and Perspectives**

Mathematical and numerical modeling of cardiovascular mechanics has been a thrilling journey of human creativity and ingenuity dating back to the work of Leonhard Euler to our days, with relevant contributions of important scientists like T.J.R. Hughes or C. Peskin. The predictive nature of mathematical and computational models has been enhancing the process of understanding the physiopathological dynamics of the cardiovascular system and of designing devices and therapeutic tools. The role of a solid mathematical and engineering background in the solution of the challenging problems arising in hemodynamics and more in general biomechanics is critical for addressing nowadays' emerging challenges. The efficiency for solving fluid-structure interaction problems, for instance, experienced terrific improvements over the last twenty years, and the solution of the complex interactions between the vascular wall and the blood flow in complex geometries is nowadays possible within reasonable timelines. This is a field where the combination of new computing architectures and new numerical methodologies has been crucial. Computational investigations are progressively becoming part not only of the medical research but also of the clinical routine in clinical trials and surgical planning. After the revolution introduced by imaging devices in the 20[th] Century, that enabled medical doctors to look "in" the patient and not just "at" the patient, mathematical and numerical models may further extend such important information to explore the future and the past of the patient. The "future" in performing upfront simulations of surgery or interventions; the "past" in providing the ground for extensive comparisons on a large scale of patients with similar clinical situations. Mathematicians, biomedical engineers, and clinical doctors are teaming up in this revolutionary experience whose final users are the healthcare and the society in general. This process is not just triggering new research and new investigations useful in other contexts, but also designing new professionals and job opportunities – as the experience of HeartFlow demonstrates. In this Chapter, we just gave basic introductory notions for presenting this complex and fascinating – intrinsically multidisciplinary – world. To complete this new revolution brought by computational mechanics a strong translational effort is required. The definition of protocols or "good simulation practices" added to the "good clinical practices" raises new challenges, ranging from (1) the efficiency of the numerical solution over large numbers of patients – where reduced-order models based on the online/offline paradigm will play a significant role – and of (2) the storing and retrieval of data - where cloud infrastructures are critical – to (3) the reliability of quantitative analyses even when the models are affected by significant uncertainty – calling for extensive data assimilation and uncertainty quantification procedures. Only a strong interdisciplinary effort can successfully give a positive answer to this challenge, in an integrated effort that impacts not only the scientific research but also the cultural background of clinical sciences, toward a progressively more established "quantitative medicine".